\documentclass[11pt,a4paper]{article}
\pdfoutput=1
\usepackage{jheppub}

\usepackage{hyperref}
\usepackage{tabularx,afterpage,xifthen}
\usepackage{amsmath,amsfonts,amssymb,mathtools}
\usepackage{slashed}
\usepackage{comment}
\usepackage{tikz}
\usepackage{caption}
\usepackage{subcaption}
\usepackage{graphicx, xcolor}

\newcommand\figref[1]{{Fig.\ \ref{#1}}}
\newcommand\secref[1]{{Sec.\ \ref{#1}}}

\newcommand{\bea}{\begin{eqnarray}}  
\newcommand{\eea}{\end{eqnarray}}

\newcommand{\nc}{\newcommand}

\nc{\beq}{\begin{equation}}
\nc{\eeq}{\end{equation}}
\nc{\barray}{\begin{eqnarray}}
\nc{\earray}{\end{eqnarray}}
\nc{\barrayn}{\begin{eqnarray*}}
\nc{\earrayn}{\end{eqnarray*}}
\nc{\bcenter}{\begin{center}}
\nc{\ecenter}{\end{center}}
\nc{\mc}{\mathcal}
\nc{\er}[1]{(\ref{eq:#1})}
\nc{\onehalf}{\frac{1}{2}} 
\nc{\partialbar}{\bar{\partial}}
\nc{\psit}{\widetilde{\psi}}
\nc{\hc}{\mbox{H.c.}}
\nc{\ev}{\;\mathrm{eV}}
\nc{\mev}{\;\mathrm{MeV}}

\nc{\gev}{\;\mathrm{GeV}}
\nc{\kev}{\;\mathrm{keV}}
\nc{\tev}{\;\mathrm{TeV}}

\newcommand{\cref}[1]{Chapter~\ref{#1}}

\def\fd{f_{D}}
\def\deltaN{\Delta N_{D}}
\def\mpd{m_{p_{D}}}
\def\med{m_{e_{D}}}
\def\alphad{\alpha_{D}}
\def\mHd{m_{H_{D}}}
\def\bd{B_{D}}

\def\lcdm{\Lambda \text{CDM}}

\title{
Precision Cosmological Constraints on Atomic Dark Matter 
}

\author{Saurabh Bansal,$^{a}$}
\emailAdd{saurabhbansal20@gmail.com}

\author{Jared Barron,$^{b}$}
\emailAdd{jared.barron@mail.utoronto.ca}

\author{David Curtin,$^{b}$}
\emailAdd{dcurtin@physics.utoronto.ca}

\author{Yuhsin Tsai,$^{c}$}
\emailAdd{ytsai3@nd.edu}

\affiliation{$^{a}$ Department of Physics, University of Cincinnati, Cincinnati, Ohio 45221,USA}
\affiliation{$^{b}$ Department of Physics, University of Toronto, Toronto, Ontario M5S 1A7, Canada}
\affiliation{$^{c}$ Department of Physics, University of Notre Dame, IN 46556, USA}

\abstract{
Atomic dark matter is a simple but highly theoretically motivated possibility for an interacting dark sector that could constitute some or all of dark matter.
We perform a comprehensive study of precision cosmological observables on minimal atomic dark matter, exploring for the first time the full parameter space of dark QED coupling and dark electron and proton masses $(\alphad, \med, \mpd)$ as well as the two cosmological parameters of aDM mass fraction $f_D$ and temperature ratio $\xi$ at time of SM recombination. 
We also show how aDM can accommodate the $(H_0, S_8)$ tension from late-time measurements, leading to a better fit than $\Lambda$CDM or $\Lambda$CDM + dark radiation. 
Furthermore, including late-time measurements leads to closed contours of preferred $\xi$ and dark hydrogen binding energy. 
The dark proton mass is seemingly unconstrained.
Our results serve as an important new jumping-off point for future precision studies of atomic dark matter at non-linear and smaller scales.
}

\begin{document}

\maketitle

\section{Introduction}

The nature of dark matter (DM) is one of the foremost mysteries of modern particle physics. All attempts to identify it through direct detection have been stymied by negative results. Model-building efforts therefore focus on explaining the production, stability, and lack of detection of dark matter so far, and exploring the possible intricacies of the dark sector. The only hints about dark matter's characteristics come from cosmological and astrophysical measurements, which largely paint a picture of dark matter that acts like a cold, collisionless fluid, leading to the primacy of the standard $\lcdm$ cosmological model. 
However, dark matter could be much less minimal while still satisfying those constraints. There are theoretical motivations for a non-minimal dark sector, including some solutions to the hierarchy problem, as we discuss below. 
Furthermore, as we have entered the era of precision cosmology, low-redshift measurements of some quantities affected by the properties of dark matter have come into tension with the values inferred from high-redshift cosmological observations within the $\lcdm$ model.
All of this makes the exploration of alternatives to minimal $\lcdm$ a high priority. 

The Hubble constant, measured by the SH0ES collaboration to be $73.04$ $\pm$ $1.04$ km/s/Mpc~\cite{Riess:2021jrx}, 
is in conflict with its inferred value from the Planck 2018 measurements of the Cosmic Microwave Background (CMB) power spectrum, $67.4 \pm 0.5$ km/s/Mpc \cite{Planck:2018vyg}. This difference constitutes a $>5 \sigma$ tension. There is also tension in measurements of the amount of large-scale structure in the universe. The tension is characterized through the quantity $\sigma_{8}$, which is defined as the root mean square of fluctuations in the matter density contrast at the scale $8 h^{-1}$ Mpc \cite{Hildebrandt:2018yau}. Often, the value $S_{8} = \sigma_{8}\sqrt{\Omega_{m}/0.3}$, where $\Omega_{m}$ is the total matter abundance, is used instead to parametrize the tension. The value predicted in the $\lcdm$ model from Planck measurements is $2-3\sigma$ higher than the value measured in various galaxy clustering and cosmic shear surveys \cite{MacCrann:2014wfa,Joudaki:2016mvz,Hildebrandt:2018yau,Joudaki:2019pmv,KiDS:2020suj,KiDS:2021opn,Heymans:2020gsg,DES:2021wwk,Philcox:2021kcw,Abdalla:2022yfr}. The statistical significances of these tensions vary, but as time passes without a resolution attributable to misunderstood systematics or errors in the local measurements, solutions to these anomalies based on physics in the dark matter sector have garnered increased attention (see e.g.~\cite{DiValentino:2020zio,Asadi:2022njl} for a general overview).

In this paper we consider the class of dark sector model known as atomic dark matter (aDM), which has strong theoretical motivations tied to symmetries and naturalness, distinctive phenomenology from cosmological down to astrophysical scales, and has been recognized as having the potential to alleviate the aforementioned tensions \cite{Kaplan:2009de,Kaplan:2011yj,Cline:2012is,Cyr-Racine:2012tfp,Cyr-Racine:2013fsa,Fan:2013tia,Fan:2013yva,Curtin:2020tkm,Bansal:2021dfh, Cyr-Racine:2021oal, Blinov:2021mdk}.

The fundamental ingredients of aDM are a hidden U(1) gauge group (which we take to be unbroken) and two fermions, neutral under the Standard Model (SM) gauge groups and oppositely charged under the hidden U(1). The hidden U$(1)$ photon and the fermions can form a tightly coupled plasma in the early universe, before the fermions combine into a hydrogen-like bound state as the universe cools. 
The mass fraction of aDM and its temperature during SM recombination  are taken to be free parameters. This defines the 5D parameter space of the minimal aDM model, with three microphysical parameters $(\alphad, \med, \mpd)$ and two cosmological parameters $(\fd, \Delta N_D)$. 

A common way to address the Hubble tension is to introduce a species of dark radiation (DR) to increase the Hubble rate at the time of last scattering, at the cost of increasing the effective number of neutrinos $N_{\rm{eff}}$ beyond the Standard Model value of 3.044 \cite{Gariazzo:2019gyi}. However, adding dark radiation alone to $\lcdm$ leads to higher $S_{8}$, worsening that tension. Adding interactions between the dark radiation and dark matter can lead to reduced growth of structure in the early universe, addressing the $S_{8}$ tension~\cite{Chacko:2016kgg,Chacko:2018vss,Rubira:2022xhb}. The atomic dark matter model contains both of these ingredients. Atomic dark matter and its cosmological signatures have been studied before~\cite{Cyr-Racine:2012tfp,Cyr-Racine:2013fsa}, but new, more precise observational datasets and the emergence of the abovementioned tensions necessitate an updated analysis. Furthermore, the full five-dimensional parameter space of even the minimal atomic dark matter model has not been fully explored. 

The subject of this paper is therefore an analysis of precision cosmological constraints on atomic dark matter, including Planck measurements of the CMB temperature and polarization spectra as well as CMB lensing~\cite{Planck:2019nip}, measurements of the baryon acoustic oscillation (BAO) feature~\cite{2011MNRAS.416.3017B,Ross:2014qpa,BOSS:2016wmc}, and Type Ia supernova lightcurves~\cite{Pan-STARRS1:2017jku}.
We also assess the model's capacity to alleviate the $H_{0}$ and $S_{8}$ tensions. 

To make precise predictions of the CMB and matter power spectrum for comparison with data, we augmented the cosmological Boltzmann-solving code CLASS \cite{2011JCAP...07..034B} to solve for the thermal history of the atomic dark matter sector, and include its effects on the evolution of density perturbations in the early universe through the Effective THeory Of Structure (ETHOS) formalism \cite{Cyr-Racine:2015ihg}. For model parameter values far from the Standard Model equivalents, an accurate computation of the dark recombination and decoupling requires significant modifications to the standard treatment in CLASS. 
We obtain constraints on the atomic dark matter parameter space by performing Markov Chain Monte Carlo scans of the posterior distribution of the model parameters given various cosmological datasets, using the MontePython code package \cite{Brinckmann:2018cvx}.

We find that cosmological constraints on aDM without the late-time $(H_0, S_8)$ measurements are modest, but still significant. The primary constraint is on the scale at which the atomic dark matter ceases to be dragged by the dark radiation. For an aDM fraction of $f_{D}=5\%$, the sound horizon at the time of the dark drag epoch is constrained to be $r_{\text{DAO}}\lesssim 10$ Mpc. If $r_{\text{DAO}} \lesssim 3$ Mpc, $f_{D}$ up to unity is allowed by CMB data. Because this scale is dominantly determined by the redshift of the dark recombination, this requirement restricts how low $\med$ and $\alphad$ can be simultaneously, and how high the dark sector temperature can be. Regardless of $r_{\text{DAO}}$, $\deltaN$ needs to obey the usual constraint $\deltaN \lesssim 0.3$ from CMB measurements~\cite{Planck:2018vyg}.
Including late-time measurements demonstrates that aDM accommodates the $(H_0, S_8)$ tension better than $\Lambda$CDM with or without dark radiation. 
To accommodate the late-time measurements, the aDM parameters conspire such that dark recombination occurs around $z\sim 3\times 10^{4}$. Additionally, $\deltaN \sim 0.3$ is preferred. Our work therefore motivates further investigation of aDM signatures at later times and smaller scales in this region of parameter space.

Our paper is organized as follows. In Section \ref{sec:ADM} we review the minimal atomic dark matter model we study, alongside its theoretical motivation and current constraints. In Section \ref{sec:cosmohistory} we describe the effects of aDM on cosmological history and cosmological observables compared to $\Lambda$CDM. Section~\ref{s.class} presents our modified CLASS code which includes aDM. Finally, Section~\ref{sec:cosmoconstraints} outlines the datasets we use to compare to computed cosmological observables and show the new constraints on the aDM parameters as well as the extent to which the model can alleviate the Hubble and $S_{8}$ tensions. We conclude in Section~\ref{sec:conclusions}.

\section{Atomic Dark Matter Review}
\label{sec:ADM}

In this section we define our simplified minimal model of atomic dark matter, briefly review how this scenario arises in more complete theories, and discuss some existing constraints on the model.

\subsection{Minimal Simplified Model}
\label{s.review}
We consider the simplest model of atomic dark matter, where some fraction of the total dark matter content of the universe is composed of fermions charged under a hidden U(1) gauge symmetry with a massless gauge boson. In order to form dark atoms, the dark sector must contain at least two fermions with equal and opposite charge. We will refer to the heavier state as a dark proton $p_{D}$, though we make no assumptions about its internal structure, and the lighter as a dark electron $e_{D}$. Due to their interaction with the $U(1)_{D}$ gauge boson, which we identify as a dark photon and assume to be massless, the dark proton and dark electron can form a bound state which we name dark hydrogen.
The model can thus be described by the Lagrangian 
\begin{equation}
\label{eq:Lagrangian}
    \mathcal{L}_{aDM} = -\frac{1}{4}A_{\mu\nu}A^{\mu\nu} + i\bar{p}_{D}(\slashed{D}-\mpd)p_{D} + i\bar{e}_{D}(\slashed{D}-\med)e_{D}
\end{equation}
where  $D_{\mu} = \partial_{\mu} + i\tilde{e}A_{\mu}$, and we define a dark fine structure constant $\alphad \equiv \tilde{e}^{2}/4\pi$. 

 We make no assumptions about the production mechanism of the atomic dark matter, other than requiring an asymmetric relic abundance and assuming that the energy density of the aDM sector is dominated by the above particle content.\footnote{If the dark proton is a bound state of some dark QCD, then additional dark nuclei might exist, analogous to helium and heavier elements. In the special case of the Mirror Twin Higgs this was studied in~\cite{Bansal:2021dfh}. We defer a fully general analysis of this possibility to future work.} We assume the dark sector only has gravitational coupling to the Standard Model in the cosmological study. In order to accommodate the high degree of consistency of a wide variety of observations with the $\lcdm$ paradigm while recognizing that a subdominant of dark matter could have non-trivial self-interactions, we define a free parameter $\fd$ as the fraction of dark matter energy density in the atomic dark matter sector, $\fd \equiv \Omega_{aDM}/\Omega_{DM}$, with the remaining dark matter being cold and collisionless. To comply with bounds on $\Delta N_{\rm{eff}}$, we allow the temperature ratio of the dark and Standard Model sectors at SM recombination, $\xi \equiv T_{D}/T_{SM}$ to be a free parameter of the model. The dark sector can naturally be colder than the visible sector if it comes from a less efficient reheating process than the visible sector, or if more species became non-relativistic and annihilated in the visible sector than the dark sector after decoupling. Several mechanisms for cooling the dark sector, including through an asymmetric reheating, have been discussed in the literature~ \cite{Berezhiani:1995am,Chacko:2016hvu,Craig:2016lyx,Beauchesne:2021opx}.
 
 By assuming that there are no other relativistic species in the dark sector, we can make an equivalence between $\xi$ and $\deltaN$, the aDM contribution to $\Delta N_{\rm{eff}}$:
\begin{equation}
\label{eq:deltaN}
    \deltaN = \left(\frac{8}{7}\right)\left(\frac{11}{4}\right)^{4/3}\xi^{4}
    \ \approx \ 4.4 \ \xi^4
\end{equation}

Our minimal simplified model of atomic dark matter thus has five free parameters. Three parameters describe the microphysics of atomic dark matter relevant to our analysis:
\begin{itemize}
\item the dark proton mass $\mpd$,
\item the dark electron mass $\med$,
\item the dark fine structure constant $\alphad$. 
\end{itemize}
Two further parameters describe the cosmological initial conditions of atomic dark matter for the purpose of determining precision cosmological observables:  
\begin{itemize}
\item    the fraction of total dark matter that is composed of atomic dark matter, $0 \leq \fd \leq 1$,
\item the dark to visible temperature ratio today, taken to be $0\leq \xi\leq 1$ (or equivalently $\deltaN$).
\end{itemize}

\subsection{Embedding in motivated Theory Frameworks}

Atomic dark matter as defined above is a simple dark sector theory constructed from a small number of fields and interactions that all have close SM analogues, making it a perfectly plausible model of dark matter in its own right, especially once it is completed by adding one of many possibilities for producing an asymmetric relic abundance in the early universe (see e.g.~\cite{Zurek:2013wia} for a review). 
However, it is important to note that aDM also arises in theory frameworks that are highly motivated for orthogonal reasons. 

Many theories consider the possibility that the SM is related to the dark sector by a discrete symmetry~\cite{Chacko:2005pe,Barbieri:2005ri,Chacko:2005vw,Chacko:2016hvu,Craig:2016lyx,Chacko:2018vss,Chacko:2021vin,GarciaGarcia:2015pnn,Bansal:2021dfh,Foot:2002iy,Foot:2003jt,Berezhiani:2003xm,Foot:1999hm,Foot:2000vy,Foot:2003eq, Foot:2004pa, Foot:2014uba, Mohapatra:2017lqw}. 
This frequently gives rise to atomic dark matter scenarios, either exact implementations of the above minimal model or generalizations that include mirror neutrinos, or heavier dark nuclei. 
For example, the so-called mirror-matter hypothesis assumes that dark matter is an exact $\mathbb{Z}_2$ copy of the SM, with many interesting cosmological and astrophysical consequences~\cite{Foot:2002iy,Foot:2003jt,Berezhiani:2003xm,Foot:1999hm,Foot:2000vy,Foot:2003eq, Foot:2004pa, Foot:2014uba, Mohapatra:2017lqw}.
One of the perhaps most theoretically motivated possibilities are models of ``neutral naturalness'' such as the Twin Higgs~\cite{Chacko:2005pe,Barbieri:2005ri,Chacko:2005vw,Chacko:2016hvu,Craig:2016lyx,Chacko:2018vss,Chacko:2021vin,GarciaGarcia:2015pnn,Bansal:2021dfh}, that solve the little hierarchy problem by introducing a dark sector related to the SM by a softly broken $\mathbb{Z}_2$ which features a dark Higgs with a modestly larger vacuum expectation value than the SM Higgs $f \sim (3 - 7) \times v$.
The result is a particular realization of atomic dark matter making up some fraction of dark matter, with twin neutrinos, dark helium, $\alphad = \alpha_{QED}$ but somewhat different masses and relative abundances than in the SM sector, which can generate a variety of cosmological signatures and alleviate the $(H_0, S_8)$ tensions~\cite{Chacko:2018vss,Bansal:2021dfh}. 
However, only the cosmology of the perfect Mirror Twin Higgs has been studied in detail~\cite{Bansal:2021dfh}, and more general implementations of the Twin Higgs framework could realize different parts of aDM parameter space. 
This connection to the hierarchy problem adds important motivation to our general study of aDM precision cosmological signatures. 

\subsection{Existing Constraints}

Compared to $\Lambda$CDM, atomic dark matter modifies cosmology and astrophysics in many ways and at many scales. In early universe cosmology, the dominant effects are additional dark radiation  from the dark photon (raising $N_{\rm{eff}}$) and dark acoustic oscillations (DAO), modifying the matter power spectrum~\cite{Cyr-Racine:2012tfp}.
The most recent directly applicable study of aDM in this context was performed almost a decade ago~\cite{Cyr-Racine:2013fsa} (see also~\cite{Archidiacono:2019wdp}). That analysis combined the microphysical parameters into a single parameter 
$\Sigma_{DAO} \equiv \alphad (B_{D}/\ev)^{-1}(m_{H_{D}}/\gev)^{-1/6}$, where $\bd=\alphad^{2} \med/2$ is the dark hydrogen binding energy and $m_{H_{D}}$ is the dark hydrogen mass, which determines the interaction rate between dark radiation and the interacting dark matter. The authors showed that for strong DM-DR interaction, the fraction of atomic dark matter has to be less than $\sim 5\%$, but this bound rapidly disappears when $\Sigma_{DAO} < 10^{-4}$, corresponding to $\med > 3 \mev$ for SM-like $\alpha_{D}$ and $\mpd$. 
Our paper extends and updates this analysis, which is motivated for several reasons. 
First is simply to update constraints with the latest cosmological data, which has greatly improved in precision in the last decade. 
Second is to understand the constraints in the full aDM parameter space, including the three microphysical parameters $(\alphad, \med, \mpd)$.
This is particularly important when the DM-DR interaction rate is not very large (which includes SM-like values), in which case the atomic dark matter fraction $f_D$ could be much closer to unity,
as demonstrated by~\cite{Bansal:2021dfh} for the subset of aDM parameter space spanned by the Mirror Twin Higgs.
An updated treatment also allows us to assess to what extent aDM in full generality could address the $(H_0, S_8)$ tension.

Going beyond linear early universe cosmology, aDM has a myriad of potentially dramatic effects spanning from galaxy cluster to stellar scales. 
However, due to the impracticality of quickly evaluating the non-linear evolution of atomic dark matter, none of them are currently appropriate for inclusion in this analysis. We provide a brief overview of them here, referring the reader to several recent reviews for more information~\cite{Cline:2021itd,Asadi:2022njl,Bechtol:2022koa,Dvorkin:2022jyg,Adhikari:2022sbh}.

On small cosmological scales, measurements of the Lyman-$\alpha$ forest and upcoming measurements of the cosmological 21-cm signal are sensitive to modifications to the matter power spectrum, and have been shown to be capable of constraining DAOs and other interacting DM scenarios~\cite{Krall:2017xcw,Bose:2018juc,Garny:2018byk,Murgia:2018now,Archidiacono:2019wdp,Munoz:2020mue,Schneider:2018xba,Lopez-Honorez:2018ipk,Escudero:2018thh}.
The astrophysical signatures of aDM at cluster and galactic scales manifest as modifications to the halo mass function, halo shape, or the formation of dark disks~\cite{Cline:2013pca,Fan:2013tia,Fan:2013yva,McCullough:2013jma,Foot:2013lxa,Foot:2013vna,Randall:2014kta,Foot:2014uba,Schutz:2017tfp,Foot:2015mqa,Chashchina:2016wle,Foot:2016wvj,Foot:2017dgx,Foot:2018dhy,Buch:2018qdr,Winch:2020cju,2021A&A...653A..86W,Kramer:2016dqu,Kramer:2016dew,Chacko:2021vin,Spergel:1999mh,Peter:2012jh,Rocha:2012jg}, due to the dissipative and self-interacting nature of aDM.
At smaller scales, aDM can form exotic objects like dark or mirror stars~\cite{Curtin:2020tkm,Winch:2020cju,Curtin:2019lhm,Curtin:2019ngc,Hippert:2021fch,Howe:2021neq,Hippert:2022snq}, or black holes with masses that cannot be generated by SM astrophysics~\cite{Pollack:2014rja,Shandera:2018xkn,Singh:2020wiq}, both of which could be detected in gravitational wave observatories. 
Additional signatures are generated if the dark photon and the SM hypercharge gauge boson have a small kinetic mixing. 
This includes visible signatures for mirror stars that accumulate SM matter from the interstellar medium~~\cite{Curtin:2019lhm,Curtin:2019ngc}, cooling of white dwarfs through accumulation of aDM and emission of dark photons~\cite{Curtin:2020tkm}, and direct detection~\cite{Chacko:2021vin,Foot:2014mia}. 

While existing observations are sensitive to these effects, connections between aDM parameters and small-scale observables have not yet been formulated with sufficient precision and generality to place constraints on the full aDM parameter space with these datasets, particularly for a subdominant aDM fraction. Accurately computing the non-linear evolution of aDM and its effects down to low redshifts and small scales requires precise n-body Magnetohydrodynamic (MHD) simulations. Fortunately, several efforts are underway to make progress towards this goal, which will benefit from the cosmological constraints on aDM parameter space we derive in our analysis. This also motivates public dissemination of our modified CLASS-aDM code (Section \ref{s.class}), since it can supply the initial conditions for these detailed baryon + CDM + aDM simulations.

\section{Cosmological History}
\label{sec:cosmohistory}
In this section we review the qualitative features of  cosmological history with aDM. This includes a detailed discussion on the thermal evolution of the dark sector, dark acoustic oscillations and structure formation. While much of this is familiar from the SM, different corners of the aDM parameter space lead to interesting new phenomena. 

\subsection{Dark Recombination and Decoupling}
At temperatures much higher than the binding energy of the dark hydrogen, the dark photons, dark electrons, and dark protons form a tightly coupled plasma in thermal equilibrium, much like the Standard Model plasma before recombination. After the temperature drops below the dark hydrogen binding energy, it becomes energetically preferable for dark protons and electrons to combine into atomic dark hydrogen. The abundance of free dark electrons begins to fall exponentially while the dark photon-baryon plasma is in equilibrium, obeying the Saha equation     \begin{equation}
\label{eq:Saha}
\frac{x_{D}^{2}}{1-x_{D}} = \frac{1}{n_{D}} \left(\frac{T_{DM} \med}{2\pi}\right)^{3/2} e^{-B_{D}/T_{DM}} ,
\end{equation}
until the dark photons decouple from the atomic dark matter, at which point the free dark electron fraction $x_{D} = n_{e_{D}}^{free}/n_{D}^{total}$ freezes out, where $n_{D}$ is the number density of all dark protons and dark hydrogen combined. $T_{DM}$ is the atomic dark matter temperature.

The full equation governing the ionization fraction $x_{D}$ for Case-B recombination, with negligible net recombination directly to the ground state, is    
\begin{equation}
\label{eq:recombination}
\dot{x}_{D} = \sum_{\ell = s,p} (x_{2\ell}\mathcal{B}_{2\ell}-n_{D}x_{D}^{2}\mathcal{A}_{2\ell})
\end{equation}
where $\mathcal{A}_{2\ell}$ and $\mathcal{B}_{2\ell}$ are the effective recombination and ionization rates to and from the $2\ell$ state of the atom, and $x_{2\ell}$ is the fraction of dark electrons occupying the $2\ell$ state \cite{1968ApJ...153....1P,1969JETP...28..146Z,Seager:1999km,Lee:2020obi}. Practically, this equation is solved in an effective multi-level formalism that accounts for the transitions between intermediate, higher energy-level states \cite{2010PhRvD..82f3521A}. 

After decoupling the dark photons begin streaming freely. If the atomic dark matter parameters are similar to the Standard Model couplings and masses, this story proceeds similarly to the Standard Model recombination. However, due to the broad range of dark sector parameters we consider, the processes controlling the thermal and kinetic decoupling of the dark photons from the dark matter can differ significantly from the SM case, and many of the assumptions used for SM recombination calculations need to be checked. 

If the dark fine structure constant $\alphad$ is small enough or the atomic dark matter number-density low enough, it is possible for the dark photons to decouple from the dark electrons and protons before they have a chance to form dark hydrogen bound states, resulting in a large fraction of dark electrons remaining uncombined. This happens when~\cite{Cyr-Racine:2012tfp} 
\begin{equation}
\label{eq:chargedbound}
\left(\frac{\alphad}{0.0073}\right)^{6}\left(\frac{\xi}{0.5}\right)^{-1}\left(\frac{\fd{} \Omega_{DM} h^{2}}{0.11}\right)\left(\frac{\mHd}{\gev} \right)^{-1} \left(\frac{\bd}{\kev}\right)^{-1} \lesssim 5.0 \times 10^{-4}\,.
\end{equation}
This corresponds to a hidden charged dark matter scenario, in which self-interaction constraints are much stronger than for a dark sector which is neutral at late times. This corner of parameter space is not our focus, since it is likely highly constrained, but the relevant physics is accurately captured in our CLASS implementation for the parameter ranges we consider.

In the Standard Model, the process keeping photons and matter in thermal equilibrium until recombination is Thomson scattering between the photons and electrons. Decoupling thus occurs roughly when the rate of energy exchange through Thomson scattering falls below the Hubble rate. For an atomic dark matter sector with particle masses and couplings that are far from Standard Model values, this is not always the case. 
If the dark sector is cold enough or has small enough $\alphad$, photo-recombination cooling, photo-ionization heating, and Bremsstrahlung heating/cooling can dominate over Thomson scattering, keeping the dark photons and dark matter in thermal contact for longer than expected from Thomson scattering alone. For parameter values where the freezeout dark electron fraction is extremely low and the ratio $\med/\mpd$ is moderate, Rayleigh scattering of dark photons with dark atoms can maintain thermal equilibrium. These effects need to be taken into account when computing the thermal history of the dark sector. The rates of these various processes are given by Eqs. \ref{eq:compton} - \ref{eq:rayleigh} below~\cite{Seager:1999km,Cyr-Racine:2012tfp}.

\begin{equation}
\label{eq:compton}
\Gamma_{T} = \frac{64\pi^{3}\alphad^{2} T_{D}^{4}}{135\med^{3}} \frac{x_{D}}{1+x_{D}}\left(1 + \left(\frac{\med}{\mpd}\right)\right)^{3}
\end{equation}

\begin{equation}
\label{eq:bremsstrahlung}
\Pi_{ff} \simeq \frac{16\alphad^{3}\bar{g}_{ff}\sqrt{2\pi T_{DM}}x_{D}^{2}n_{D}^{2}}{(3\mu_{D})^{3/2}} \left(\frac{\pi^{2}\epsilon (1+2\epsilon) - 6\zeta(3)\epsilon^{2}}{6}\right)
\end{equation}
\begin{equation}
\label{eq:photorec}
\Pi_{\rm{p-r}} = \frac{2\alpha_{D}^{3} \sqrt{2\pi T_{DM}}}{3\mu_{D}^{3/2}} x_{D}^{2}n_{D}^{2}F_{\rm{p-r}}(\frac{T_{D}}{B_{D}},\frac{T_{DM}}{T_{D}})
\end{equation}
\begin{equation}
\label{eq:photoion}
\Pi_{\rm{p-i}} = \frac{\alpha_{D}^{3} T_{D}^{2}}{3\pi}x_{2s}n_{D}e^{-\frac{B_{D}}{4T_{D}}}F_{\rm{p-i}}(T_{D}/B_{D})
\end{equation}
\begin{equation}
\label{eq:rayleigh}
\Pi_{R} \simeq \frac{430080\zeta(9)\alphad^{2}n_{D} (1-x_{D})T_{D}^{9}\epsilon}{\pi^{2}\bd^{4}\mHd\med}\,.
\end{equation}
The $\Pi_{i}$ are the volumetric rates for each of these processes - Bremsstrahlung, photo-recombination, photo-ionization, and Rayleigh scattering, respectively. $T_{D}$ is the dark photon temperature, $T_{DM}$ is the atomic dark matter temperature, $\mu_{D}$ is the reduced mass of the dark electron, $\epsilon = \frac{T_{D}-T_{DM}}{T_{DM}}$, $\bar{g}_{ff} = 1.3$ is the Gaunt factor, $F_{\rm{p-i}}$ and $F_{\rm{p-r}}$ are numerical functions of the dark sector temperatures, and $\zeta$ is the Riemann zeta function. 

The Boltzmann equation governing the temperature evolution of the atomic dark matter is \cite{Seager:1999km,Cyr-Racine:2012tfp}
\begin{equation}
\label{Tmat_Boltzmann}
\frac{dT_{DM}}{dz} = \frac{1}{1+z}(2T_{DM} + \frac{2(\Pi_{\rm{p-r}} - \Pi_{\rm{p-i}}-\Pi_{ff} + \Pi_{R})}{3k_{B}n_{D}(1+x_{D})H(z)} + \Gamma_{T}\frac{T_{DM}-T_{D}}{H(z)})
\end{equation}
As long as any of the processes described above exchanges energy between the dark matter and dark radiation at a rate higher than Hubble, the two stay in thermal equilibrium at the same temperature. Once these processes all cease to be efficient, the dark matter temperature $T_{DM}$ starts evolving adiabatically, and decreases faster than the dark photon temperature $T_{D}$. Rayleigh scattering and photo-ionization can also affect the opacity of the dark plasma, reducing the mean free path of dark electrons and delaying kinetic decoupling.

As we discuss in Section~\ref{s.class}, these additional effects have been included in our modified CLASS code, with the exception of the photo-recombination and photo-ionization, due to computational issues. Fortunately, this omission does not significantly change the thermal evolution of the atomic dark matter, due to the parametric similarity between the Bremsstrahlung and photo-heating rates. This issue is explored in more detail in Section~\ref{s.class}.

\begin{figure}
\centering
\includegraphics[scale=0.7]{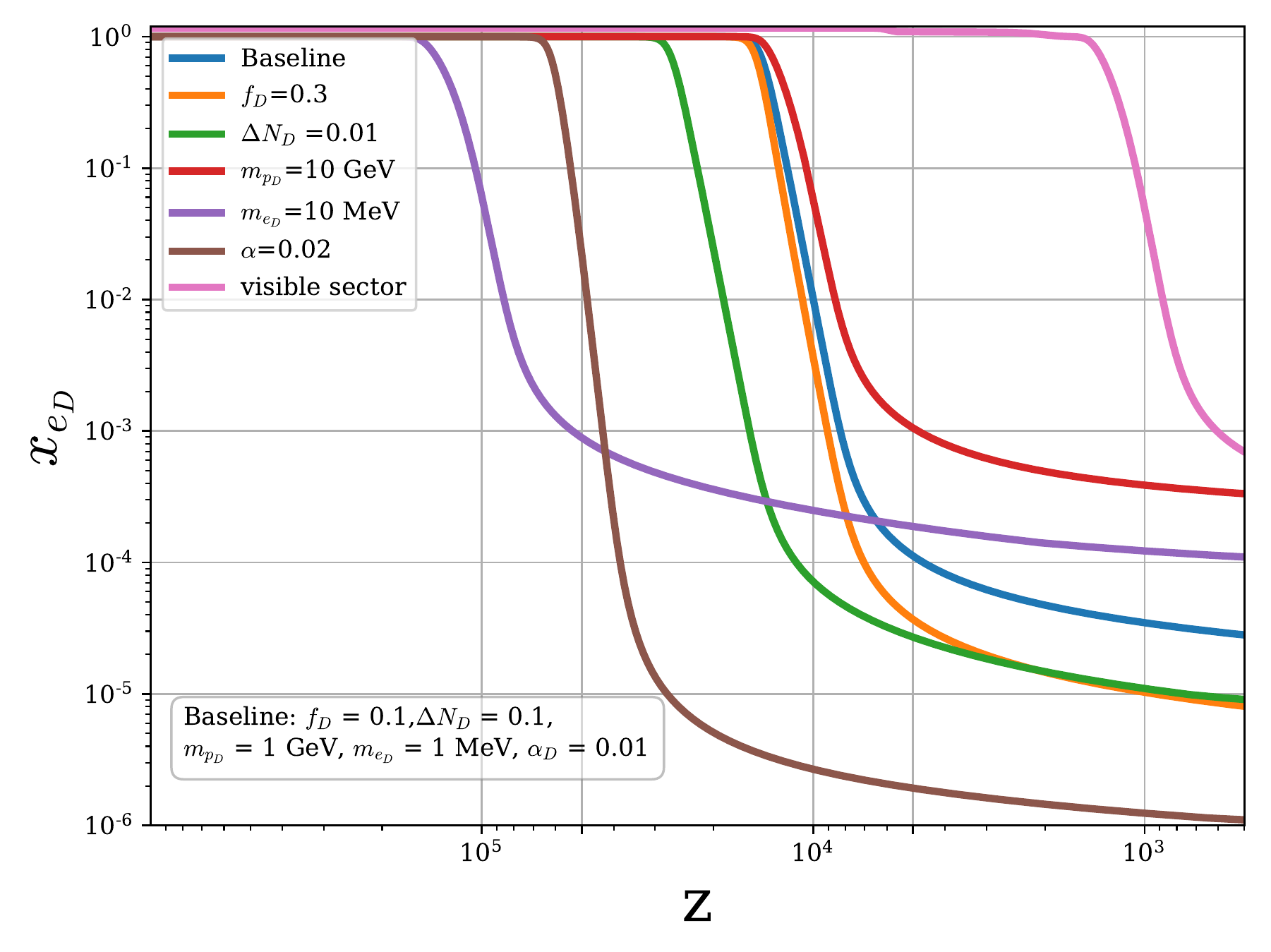}
\caption{Dark ionization fraction as a function of redshift for different parameter choices in the atomic dark matter sector.}
\label{fig:darkrecombination}
\end{figure}

To guide the reader's intuition for the dependence of the recombination history in the dark sector on the model parameters, Fig.~\ref{fig:darkrecombination} shows the evolution of the free dark ionization fraction $x_{e}$ for a variety of choices of atomic dark matter parameters. The redshift of the dark recombination is controlled primarily by
$B_{D}/\xi$. Higher binding energy and lower temperature ratio give an earlier recombination, as the green, brown, and purple curves in Fig. \ref{fig:darkrecombination} show. Higher $\mHd$, corresponding to lower atomic dark number density, leads to earlier freeze-out due to lower Thomson scattering rate, and higher residual $x_{D}$. Higher $\fd$ has the opposite effect, since the number density increases. Varying these five parameters can lead to wildly different ionization histories, with arbitrary redshift of recombination and freeze-out ionization fraction. 

\subsection{Structure formation}
Once the thermal and ionization history has been solved for, the evolution of density fluctuations in the dark sector can be determined. The Boltzmann equations governing the evolution of the dark photon perturbations, in conformal Newtonian gauge, are: 
\begin{equation}
\dot{\delta}_{\gamma_{D}} + \frac{4}{3} \theta_{\gamma_{D}} - 4\dot{\phi} = 0
\end{equation}
\begin{equation}
\dot{\theta}_{\gamma_{D}} + k^{2}(F_{\gamma_{D},2} - \frac{1}{4}\delta_{\gamma_{D}}) - k^{2}\psi = -\frac{1}{\tau_{D}}(\theta_{\gamma_{D}}-\theta_{b})
\end{equation}
\begin{equation}
\dot{F}_{\gamma_{D},\ell} = \frac{k}{2\ell + 1}((\ell+1)F_{\gamma_{D},\ell} + (1- \ell) F_{\gamma_{D},\ell-1}) = -\frac{1}{\tau_{D}}\alpha_{\ell}F_{\gamma_{D},\ell},\quad(\text{for }\ell\geq2)
\end{equation}
$\delta_{i}$ is the density fluctuation of species $i$, $\theta_{i}$ is the divergence of the velocity, and $F_{\gamma_{D}\ell}$ is the $\ell$-th moment of the dark photon temperature perturbation. $k$ is the comoving wave number of the perturbation. The factor $\alpha_{\ell}$ is related to the angular structure of the scattering cross-section between the dark photons and fermions. For atomic dark matter, $\alpha_{2} = 9/10$, $\alpha_{\ell\geq3} = 1$. $\phi$ and $\psi$ are the gravitational potentials.  
The Boltzmann equations governing the evolution of the dark proton perturbations are, following the notation of \cite{Ma:1995ey}: 
\begin{equation}
\dot{\delta}_{D} + \theta_{D} - 3 \dot{\phi} = 0
\end{equation}
\begin{equation}
\dot{\theta}_{D} + \frac{\dot{a}}{a} \theta_{D} - c^{2}_{D} k^{2} \delta_{D} - k^{2} \psi = -\frac{4\rho_{\gamma_{D}}}{3\rho_{D}}\frac{1}{\tau_{D}} (\theta_{D} - \theta_{\gamma_{D}})
\end{equation}
$\tau_{D}$ is the opacity of the dark plasma. For the Standard Model, it typically is computed including only Compton scattering. For atomic dark matter, we also include the contributions of Rayleigh scattering and photo-ionization, so that $\tau_{D}$ is defined as 
\begin{equation}
\tau_{D}^{-1} = \tau_{\text{Compton}}^{-1} + \tau_{\text{Rayleigh}}^{-1} + \tau_{\text{p-i}}^{-1}
\end{equation}
\begin{equation}
\tau_{\text{Compton}}^{-1} = an_{D}x_{D}\sigma_{T,D}\left(1 + \left(\frac{m_{e_{D}}}{m_{p_{D}}}\right)^{2}\right)
\end{equation}
\begin{equation}
\tau_{\text{Rayleigh}}^{-1} \simeq 32\pi^{4} a n_{D} (1-x_{D})\sigma_{T,D}\left(\frac{T_{D}}{B_{D}}\right)^{4},\quad\quad T_{D}<< B_{D}
\end{equation}
\begin{equation}
\tau_{\text{p-i}}^{-1} \simeq a n_{D} x_{2s} e^{-B_{D}/(4T_{D})}\frac{\sqrt{\pi}\med^{3/2}}{4\sqrt{2}\zeta(3)T_{D}^{3/2}}(\frac{\alphad}{\alpha_{SM}})^{3}\mathcal{A}_{2s}^{SM}(T_{D},T_{D})
\end{equation}
where $\sigma_{T,D}$ is the Thomson cross-section $8\pi\alpha_{D}^{2}/3m_{e_{D}}^{2}$. By numerically solving these coupled equations, we can accurately predict the evolution of density fluctuations and observe the dark acoustic oscillations imprinted on the matter power spectrum.

\subsection{Dark acoustic oscillations}

\begin{figure}
\centering
\includegraphics[scale=0.7]{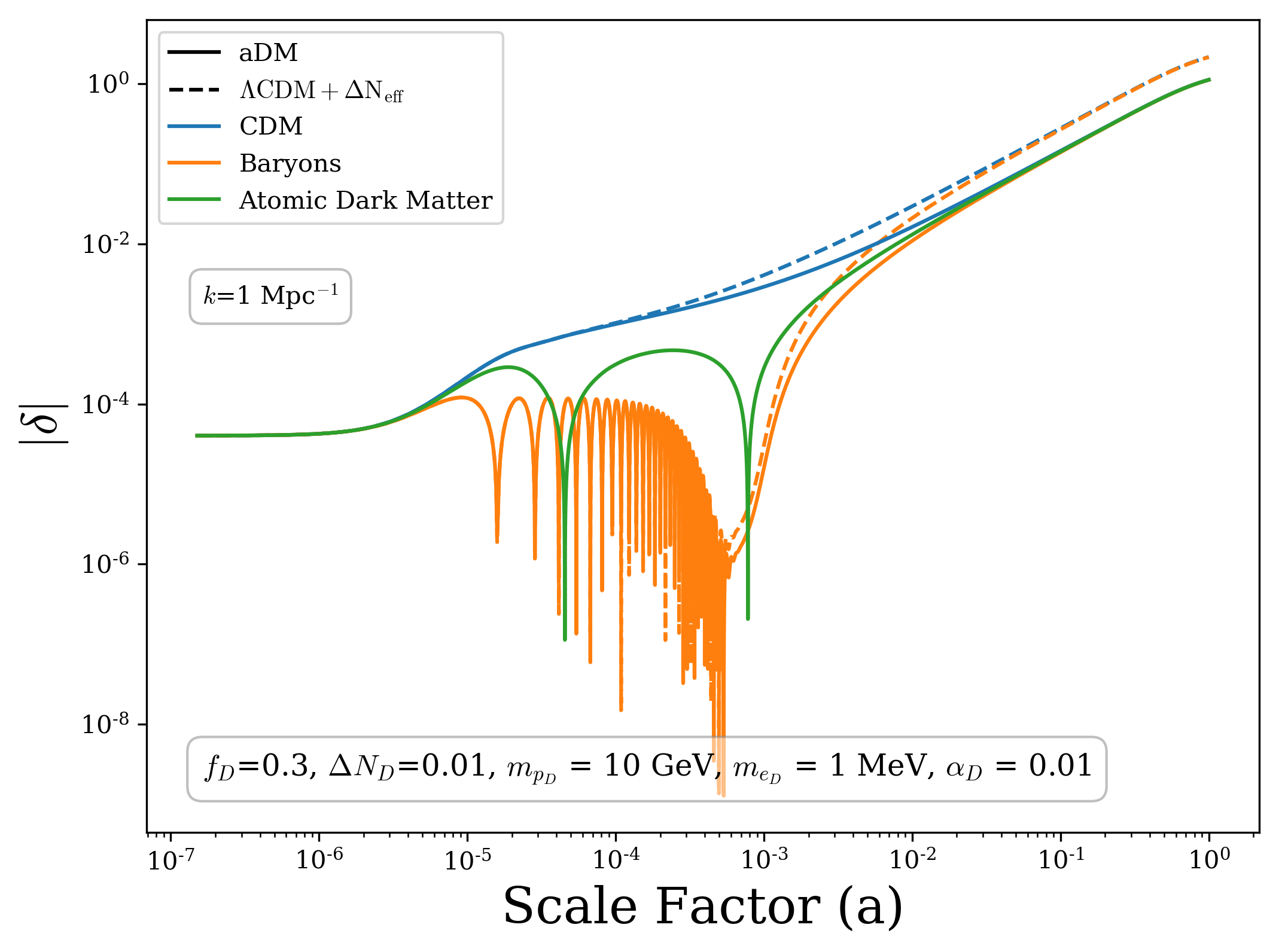}
\caption{Linear evolution of one mode for CDM, aDM, and baryons. The dashed lines show the evolution of CDM and baryons in the $\Lambda$CDM + $\Delta N_\mathrm{eff}$ model, demonstrating the suppression of structure from DAO in aDM.
}
\label{fig:onemode}
\end{figure}

Before dark recombination, if the dark photons and dark fermions interact strongly enough, they can form a tightly coupled plasma, just as the Standard Model photons, electrons, and baryons do. Perturbation modes that enter the horizon before the dark plasma decouples begin oscillating, as gravitational collapse competes with the dark photon radiation pressure. The modes stop oscillating when the dark photons decouple, and the perturbations in the atomic dark matter begin growing linearly with $a$ during matter domination, like cold dark matter. Since different modes have different phases based on when they entered the horizon, the oscillations are imprinted on the matter power spectrum, exactly as baryon acoustic oscillations are.  Also, since some fraction of dark matter is tied up in the oscillating plasma at early times, the total dark matter perturbations do not grow as quickly as in $\lcdm{}$, leading to a suppression of the matter power spectrum for modes that enter the horizon before the dark plasma decoupling. This behaviour is shown in Fig. \ref{fig:onemode}, which tracks the evolution of perturbations in the baryons, atomic dark matter, and cold dark matter at $k=1 $Mpc$^{-1}$ as a function of redshift, and compares to the evolution of perturbations in $\lcdm{}$. Both the dark acoustic oscillations and baryon acoustic oscillations are clearly visible, as is the suppression of the growth of the perturbation relative to $\lcdm{}$. 
These dark acoustic oscillations are the most distinctive signature of this model. They imprint themselves on the matter power spectrum, and can couple to the Standard Model photon temperature fluctuations through the gravitational perturbations.

\subsection{Cosmological Observables}

In this section, we discuss the features of the aDM model in the matter power spectrum and CMB. This will help us in understanding the results of Section \ref{sec:cosmoconstraints}, in which scans of the posterior distribution of the model parameters with respect to various datasets are discussed. 

\subsubsection{Large Scale Structure}
\begin{figure}
\centering
\includegraphics[scale=0.7]{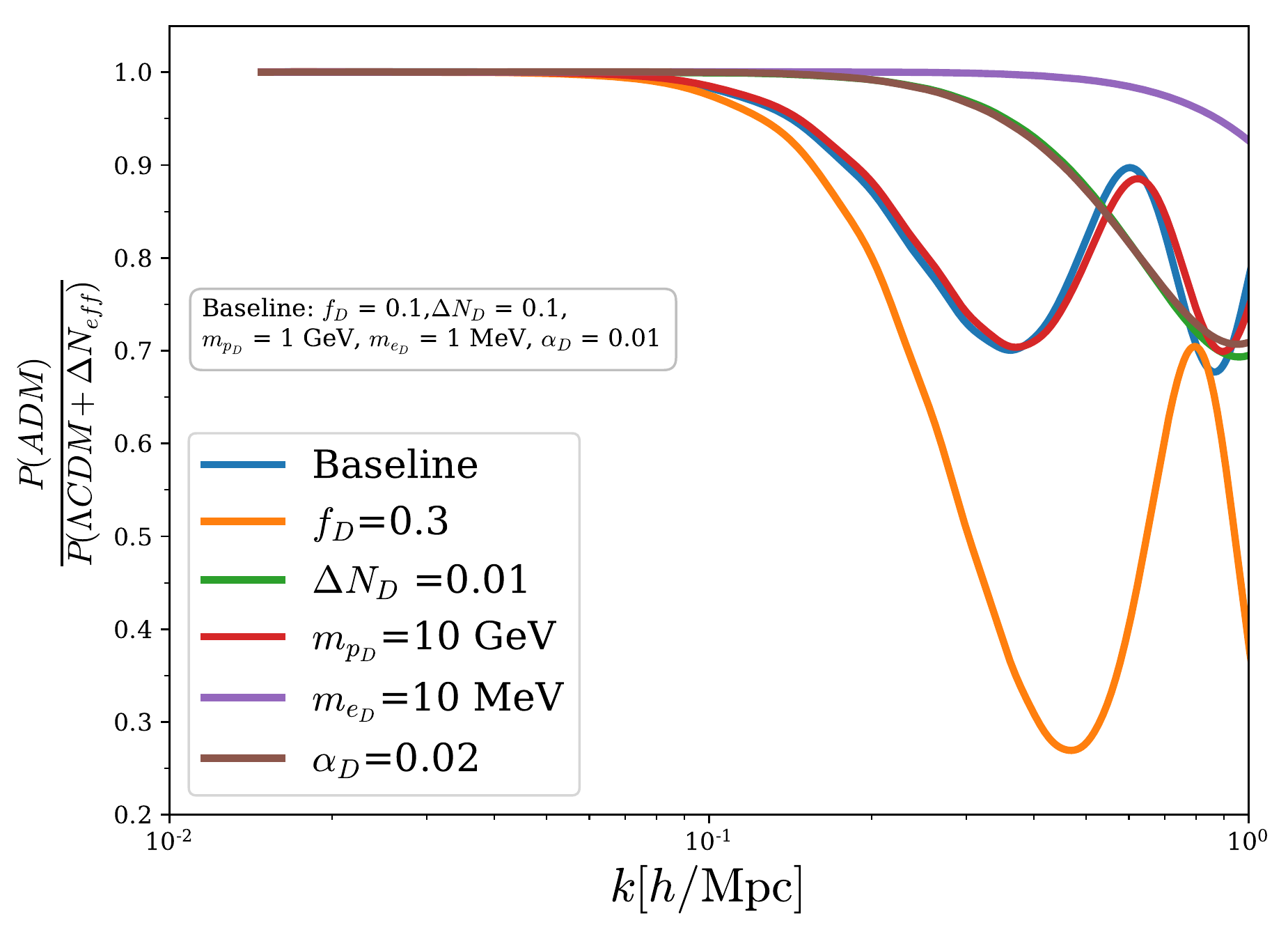}
\caption{Linear matter power spectrum in aDM relative to $\lcdm+\Delta N_{\rm{eff}}$. Dark acoustic oscillations are visible. The baseline model corresponds to $f_D$ = 0.1, $\Delta N_D$ = 0.1, $m_{p_D}$ = 1 GeV, $m_{e_D}$ = 1 MeV and $\alpha_D$ = 0.01
}
\label{fig:mpk}
\end{figure}

The direct effect of an atomic dark sector can be observed in large scale structures (LSS). 
aDM modes that enter the horizon before decoupling oscillate rather than grow. 
This leads to a suppression in the power spectrum compared to the $\lcdm$ model at scales smaller than the horizon at the decoupling epoch. 
In addition, different modes stop oscillating at different phases, leading to oscillatory features in the linear power spectrum as well. 
In Fig.~\ref{fig:mpk}, the ratio of linear matter power spectrum in the aDM model is plotted with respect to $\lcdm{}$+$\Delta N_{\rm eff}$ model.
As expected, the matter power spectrum at large $k$-modes is suppressed as compared to the $\lcdm{}$+$\Delta N_{\rm eff}$ model. 
These are the modes that enter the horizon before the dark recombination.
Note that the suppression is directly proportional to $(1-f_D)^2$ and agrees with the analytical calculations of Ref.~\cite{Chacko:2018vss}.
In addition to the overall suppression, we also see oscillations in the ratio of the matter power spectrum, which capture the phase of different $k$-modes at dark recombination.

For scales larger than $k\approx 0.1 h$~ Mpc$^{-1}$, non-linear effects mix and re-distribute perturbations among higher $k$-modes, which would wash-out the oscillatory feature in Fig.~\ref{fig:mpk}~\cite{2008ApJ...686...13S,PhysRevD.90.043524,Bernardeau:2001qr}. For instance, Ref.~\cite{Schaeffer:2021qwm} shows that the dark acoustic oscillation features in the matter power spectrum can be washed out at redshift below $z\approx 1$ using N-body simulations. However, future surveys such as the Euclid~\cite{Amendola:2016saw} mission and the LSST/Vera Rubin Observatory~\cite{LSSTDarkEnergyScience:2018jkl} can be sensitive to the matter power spectrum with $z>1$. This gives a chance to probe the oscillatory feature of the matter perturbation. Moreover, the oscillatory feature can also remain in the halo mass function~\cite{Schaeffer:2021qwm}.

In this work, we only calculate the matter power spectrum of the aDM model with linear Boltzmann equations. This gives a reasonable approximation of the CMB signals we consider since the $\ell$-modes within the Planck sensitivity mainly come from perturbations with  $k\lesssim 0.1h$~Mpc$^{-1}$. We do include cosmic shear measurements of the matter power spectrum from the KiDS+VIKING-450 (KV450) dataset in some of the MCMC studies, but in order to minimize the effect of non-linear corrections, we only use this dataset for $k\lesssim 0.2 h$~Mpc$^{-1}$. With this restriction, we find that the KV450 data does not change the constraints on aDM parameters by much, as shown later in \figref{fig:KV450_withvswithout}.      
When presenting bounds on the aDM parameters, we plot the energy density ratio all the way to $f_D=1$. If $f_D\approx1$, the dark acoustic oscillations may suppress the matter power spectrum significantly, and the result may violate bounds from the Lyman-$\alpha$ forest~\cite{Garny:2018byk} and sub-halo mass function (SHMF)~\cite{DES:2020fxi} measurements. However, the calculation of the Lyman-$\alpha$ and SHMF bounds for the aDM model is beyond the scope of this work. When $f_D\gg 0.1$, one should take our bound with caution since the scenario can be further constrained by these other observations.

\subsubsection{CMB}
A subdominant effect of the aDM sector, but one with much more constraining power, can also be observed in the CMB, which leads to less than $\mathcal{O}(1\%)$ changes in the CMB power spectrum compared to the $\lcdm{}$ model.
Though small, these changes are within the reach of the current precision of the CMB measurements by the Planck collaboration~\cite{Planck:2018vyg,Planck:2019nip}.
While the constraints on the aDM parameter space from the Planck data are discussed in the next section, here we provide qualitative arguments on the relationship between aDM and the CMB.

\begin{figure}
\centering
\includegraphics[scale=0.7]{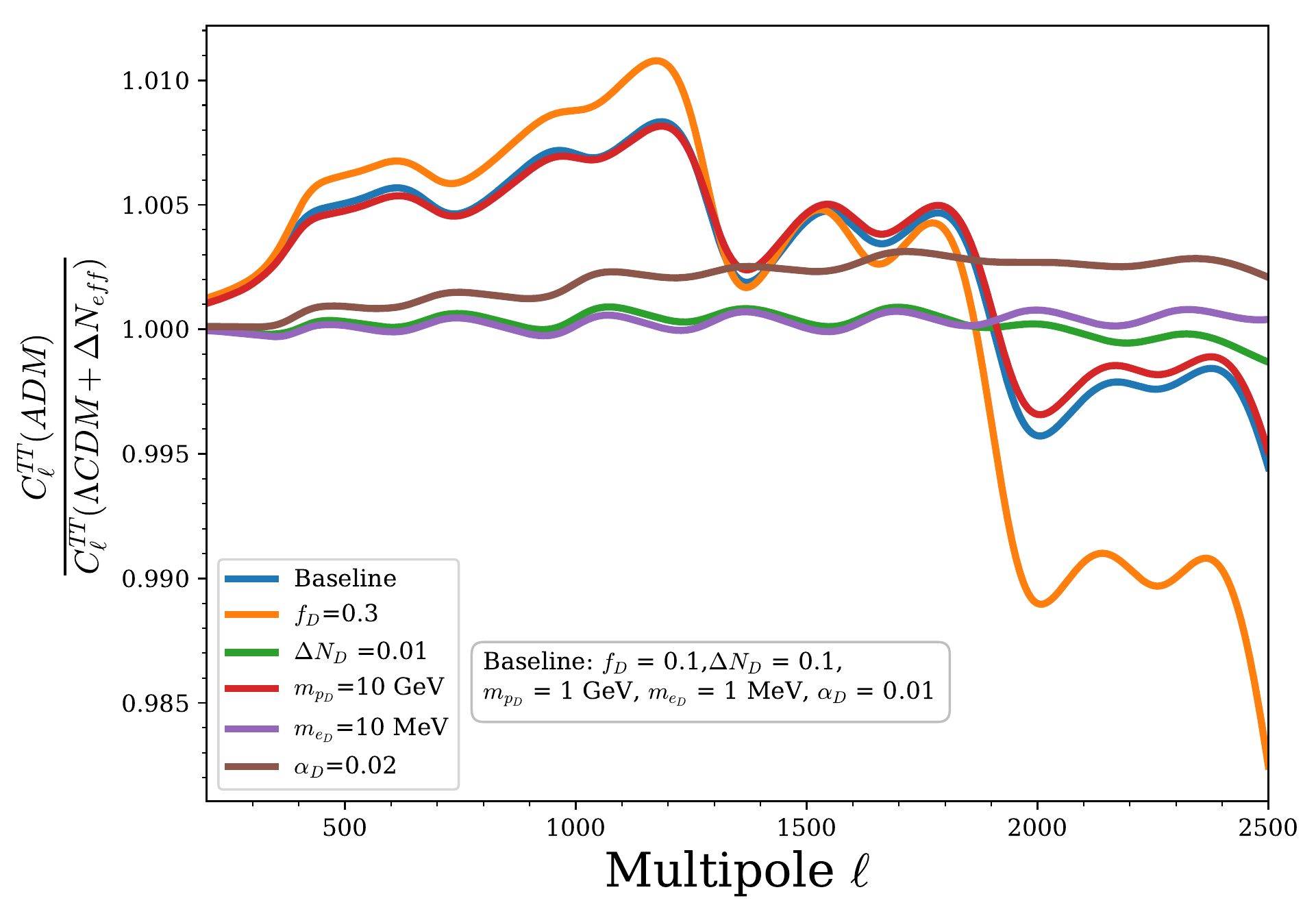}
\caption{CMB temperature power spectrum relative to $\lcdm+\Delta N_{\rm{eff}}$. As before, the baseline model corresponds to $f_D$ = 0.1, $\Delta N_D$ = 0.1, $m_{p_D}$ = 1 GeV, $m_{e_D}$ = 1 MeV and $\alpha_D$ = 0.01. 
}
\label{fig:ClTT}
\end{figure}

In Fig.~\ref{fig:ClTT}, we show the temperature anisotropies for the aDM model with respect to those in the $\lcdm{}$+$\Delta N_{\rm eff}$ model, for a variety of aDM parameters.
Note that the highest deviations in the figure are $\mathcal{O}(1\%)$ and are much smaller than those on the matter power spectrum.
These deviations are primarily due to scattering dark photons and suppressed gravity perturbations in the aDM model. Unlike free-streaming radiation in the $\Lambda$CDM+$\Delta N_\mathrm{eff}$ model, the dark photons in the aDM model are fluid-like until recombination and free stream after that. 
This leads to a phase shift in the photon density perturbations ($\delta_\gamma$) and thus, in $C_\ell^{TT}$~\cite{Bashinsky:2003tk,2013PhRvD..87h3008H,Baumann:2015rya,Chacko:2015noa}.
On the other hand, the suppressed gravity perturbation due to DAO changes the amplitude of $\delta_\gamma$ by shifting the equilibrium point of the oscillations, which is driven by the tug-of-war between the gravitation pull and radiation pressure. See Ref.~\cite{Bansal:2021dfh} for more discussions on the DAO modification of the CMB spectra.

A combination of the above mentioned effects lead to the deviations we observe in Fig.~\ref{fig:ClTT}. On increasing $f_D$, the gravity perturbations are more suppressed, leading to a bigger shift in the equilibrium point of $\delta_\gamma$ oscillations and thus, bigger deviations in $C_\ell^{TT}$. 
On the other hand, a smaller $\deltaN$, higher $m_{e_D}$ or higher $\alpha_D$ all lead to an earlier recombination, and thus, the $C_\ell^{TT}$ spectra for these cases are closer to the $\lcdm{}$+$\Delta N_{\rm eff}$ model.
Interestingly, increasing $m_{p_{D}}$ from 1 GeV to 10 GeV has minimal impact on the CMB spectrum.

\section{aDM CLASS Code}
\label{s.class}

We implemented an atomic dark matter sector in the cosmological Boltzmann code CLASS, building on a previous version which was specific to the Mirror Twin Higgs model, which is available \href{https://github.com/srbPhy/class_twin}{here}.\footnote{\texttt{https://github.com/srbPhy/class\_twin}}  The code allows for the model parameters $\fd,~\xi,~\med,~\mpd$ and $\alphad$ to be given as inputs to CLASS. Our code is public and can be downloaded \href{https://github.com/jp-barron/class\_adm-3.1.git}{here}.\footnote{\texttt{https://github.com/jp-barron/class\_adm-3.1.git}}  
We first solve the thermal and recombination history of the atomic dark matter, and use this information to compute the opacity of the dark plasma to dark radiation, which enters into the Boltzmann equations for the evolution of density fluctuations. 

The dark sector recombination is computed by initially allowing the dark ionization fraction to evolve as determined by the Saha equation Eq.~\ref{eq:Saha} instead of the full Boltzmann equation, due to the stiffness of the Boltzmann equation for the dark ionization fraction while the dark plasma is in thermodynamic equilibrium at early times. At a redshift determined by the time when the dark plasma begins to depart from equilibrium, the dark ionization fraction switches to being evolved using the full Boltzmann equation \ref{eq:recombination} in the effective multi-level atom approximation, handled by the code HyRec 2.0 \cite{2010PhRvD..82f3521A,2011PhRvD..83d3513A,Lee:2020obi}. We use an extended table of recombination coefficients with the effective number of levels taken into account equal to infinity by computing the recombination coefficients numerically   
for large $n$ values\footnote{We thank Yacine Ali-Ha\"{i}moud and Nanoom Lee for providing helpful code.} and extrapolating to infinity by fitting to the formula
\begin{equation}
\label{nlevel_extrap}
\mathcal{A}_{n}(T_{D},T_{DM}) =  \mathcal{A}_{\inf}(T_{D},T_{DM})(1-\kappa(T_{D},T_{DM})/n^{\gamma(T_{D},T_{DM})})
\end{equation}
for each temperature value. We thus are using the most accurate available recombination coefficients. This extension is necessary because the table of recombination coefficients used by HyRec does not reach to sufficiently high enough temperature/binding energy ratios to capture the recombination behaviour of the atomic dark sector for all parameter choices.

To maintain the stability of the code across the wide range of possible recombination and decoupling histories, the redshift at which the switch from Saha to Boltzmann occurs is defined as follows: if the dark photons and atomic dark matter thermally decouple before $x_{D}$ would decrease to $0.999$, the switch to Boltzmann evolution is triggered when $\Gamma_{T}$ falls below $100 H$. If the decoupling happens near the time of recombination, the switch occurs when $x_{D}=0.999$. Finally, if the dark plasma is in equilibrium long past dark recombination, the Saha equation is used until $x_{D}=10^{-7}$. These choices were tested extensively to ensure that the code accurately computes the dark ionization history even for very early or late decoupling of the dark plasma, including for the entire range of parameter values defined in section~\secref{s.planckbaopantheon} that we use in our scans.  

The thermal coupling of the dark radiation and atomic dark matter was computed using a quasi-static approximation in the high-temperature limit, as is usual in CLASS, for stability of the code when the Boltzmann equation is very stiff. However, because non-Thomson processes can dominate over Thomson scattering and keep the atomic dark matter in thermal equilibrium with the dark photon bath, the default CLASS treatment needed to be augmented. We included the effects of Rayleigh scattering and free-free (Bremsstrahlung) scattering as well, as described by \eqref{eq:rayleigh} and \eqref{eq:bremsstrahlung}. After the highest of the three scattering rates falls below $2000\times H_{0}$, the temperature evolution is calculated using the full Boltzmann equation \eqref{Tmat_Boltzmann}. This threshold was chosen empirically to maintain the stability of the code for a wide range of input parameters. The photo-recombination and photo-ionization processes were not included in the version of the code used for this analysis, due to numerical instabilities in the cancellation of the two terms for general atomic dark matter parameter values. However, we are confident that this has little effect on our results. This is because the time of thermal decoupling is determined by whether any energy-exchanging process is proceeding efficiently, not the exact rate, and the Bremsstrahlung scattering rate has the same parametric dependence as the photo-heating processes. Therefore, when these processes are dominant, the redshift where the Bremsstrahlung rate drops below Hubble is very close to the redshift where the photo-heating rates drop below Hubble. The impact of the omission of the photo-heating processes is therefore negligible. We have confirmed that our code closely reproduces the aDM temperature evolution shown in \cite{Cyr-Racine:2012tfp} for a choice of parameters where Bremsstrahlung and photo-heating control the thermal decoupling, while only including Bremsstrahlung.  

The evolution of dark matter perturbations are handled by the Effective THeory Of Structure (ETHOS) framework in CLASS. After the dark ionization fraction and atomic dark matter temperature evolution are calculated, the built-in ETHOS implementation in CLASS is used, and the atomic dark matter is treated as a sector of interacting dark matter and dark radiation. The opacity $\dot{\kappa}$ of the dark sector is set by the Thomson, Rayleigh, and photo-ionization scattering rates. The conversion from aDM parameters to those used by ETHOS is outlined in Table~\ref{tab:ETHOS}.

\begin{table}
\renewcommand*{\arraystretch}{2}
\begin{center}
\begin{tabular}{|c || c | c | c| c | c | c | c | c | c |}
\hline
ETHOS & DR & $\chi$ & $\dot{\kappa}_{DR-DM}$ & $\dot{\kappa}_{\chi}$ & $c_{\chi}^{2} $& $\alpha_{\ell = 2}$ & $\alpha_{\ell\geq 2}$ & $\dot{\kappa}_{DR-DR}$ & $\beta_{\ell}$\\ \hline
aDM & $\gamma_{D}$ & $H_{D}$ & $-\frac{1}{\tau_{D}}$ & $-\frac{4\rho_{\gamma_{D}}}{3\rho_{D}}\tau_{D}^{-1}$ & $c_{D}^{2} $& 9/10 & 1 & 0 & 0\\
\hline
\end{tabular}
\end{center}
\caption{Conversion between ETHOS and aDM parameters.}
\label{tab:ETHOS}
\end{table}

There are several assumptions underlying the usual SM recombination calculation which are also made in our aDM recombination calculation, but may break down far from SM parameter values. We include flags in the code that warn when one or more of these assumptions are violated. For very low number densities or weak coupling, violating the bound in \ref{eq.casea}, recombination can fail to be Case-B dominated, and the net rate of recombinations to the ground state can be significant, which would necessitate the use of a Case-A recombination coefficient. For an extremely cold dark radiation bath or large dark fine structure constant violating the bound in \ref{eq.nonthermaldr}, the energy injections from recombination and other processes can significantly contribute to the radiation energy density, disrupting the assumption that the dark photons are thermally distributed \cite{Cyr-Racine:2012tfp}.  Finally, collisional recombination can contribute significantly to the net recombination rate for a sufficiently cold or weakly coupled dark sector, if the bound in \ref{eq.collrec} is violated. When any of these bounds are violated, our CLASS code can no longer be trusted to accurately compute the dark recombination correctly. Including these effects in the CLASS-aDM code is left for future work. The aDM parameter ranges used in our parameter scans were chosen to respect these bounds down to at least $\fd = 10^{-3}$ and $\deltaN = 10^{-4}$.  
\begin{equation}
\label{eq.casea}
\alphad^{-6}\xi^{3/2}\fd^{-1}\left(\frac{\mpd}{\gev}\right) \left(\frac{\med}{\gev}\right)^{-1} < 2.5\times 10^{24}
\end{equation}
\begin{equation}
\label{eq.nonthermaldr}
200 \alphad^{4} \xi^{-4} \left(\frac{\mpd}{\gev} \right)^{-2}\left(\frac{\med}{\ev}\right)^{-1} \lesssim 0.1
\end{equation}
\begin{equation}
\label{eq.collrec}
\alphad \xi^{3} \left(\frac{\mpd}{\gev}\right) \fd^{-1} > 10^{-10}
\end{equation}

With these modifications, our version of CLASS can compute CMB and matter power spectra for very wide ranges of aDM parameter choices, including $\fd$ and $\deltaN$ from 0 to 1, and masses and couplings spanning many orders of magnitude, see Section~\ref{s.planckbaopantheon}. 

\section{Cosmological Constraints}

In this section we compare the predicted CMB spectrum, matter power spectrum, and Hubble rate within the atomic dark matter model with real observations in order to compare it to $\Lambda\rm{CDM}$, with and without an arbitrary amount $\Delta N_{\rm eff}$ of dark radiation, and derive the best-available current constraints on aDM model parameters. We find that aDM is quite unconstrained without late-time $H_0$ and $S_8$ measurements, but when adding these constraints we find that aDM parameters have to fall into well-defined windows.

\label{sec:cosmoconstraints}

\subsection{Datasets}
\label{s.datasets}
To evaluate the constraints on atomic dark matter from cosmological observations, we included the following experimental datasets in our analysis. 
\begin{itemize}
\item From the Planck 2018 release, we use the high-$\ell$ TTTEEE, low-$\ell$ EE, low-$\ell$ TT, and lensing datasets \cite{Planck:2019nip}. 
\item We include the measurements of the BAO feature by various galaxy surveys, reported as $D_{V}/r_{s}$ by 6dFGS at $z=0.106$\cite{2011MNRAS.416.3017B}, by SDSS in Data Release 7 at $z=0.15$\cite{Ross:2014qpa}, and by BOSS in Data Release 12 at $z=0.2 - 0.75$\cite{BOSS:2016wmc}. 
\item We also include the Pantheon supernova likelihood, which constrains the relationship between redshift and distance at low redshift. \cite{Pan-STARRS1:2017jku}.
\item  To compare the impact of atomic dark matter with direct measurements of large-scale structure, we use the KiDS+VIKING-450 cosmic shear dataset \cite{Hildebrandt:2018yau}, with a cut-off at $k = 0.2h\ \rm{Mpc}^{-1}$ to minimize exposure to the non-linear regime.  

\item To quantify the Hubble tension and examine the impact of including direct measurements of $H_{0}$ on the preferred parameter space, we use the most recent measurement of the Hubble constant using distance ladder methods by the SH0ES collaboration, $H_{0} = 73.04 \pm 1.04\  \rm{km/s/Mpc}$\cite{Riess:2021jrx}. 

\item To test the model's capacity for addressing the $S_{8}$ tension, we use the measurement from the KiDS-1000 survey, which was determined through a multi-probe analysis of cosmic shear, galaxy clustering, and galaxy-galaxy lensing \cite{Heymans:2020gsg}. This analysis found $S_{8}\equiv \sigma_{8}\sqrt{\Omega_{m}/0.3} = 0.766^{+0.020}_{-0.014}$, reported as being in $2-3\sigma$ tension with Planck.
\end{itemize}

We obtain constraints with two different combinations of datasets.
Our baseline set of datasets is the Planck, BAO, and Pantheon measurements, which we refer to as Planck+BAO+Pantheon. We use this set of experiments to set robust constraints on the atomic dark matter model parameters in Section~\ref{s.planckbaopantheon}.
To explore the impact of including large-scale structure measurements on those constraints, we add the KV450 dataset to our baseline dataset in Section~\ref{s.planckbaopantheonkv450}. We find this to have minimal impact.
To quantify the improvement that atomic dark matter can yield in the $H_{0}$ and $S_{8}$ tensions simultaneously, we add the SH0ES and KiDS-1000 measurements to the baseline datasets in Section~\ref{s.planckbaopantheonshoeskids}, and show the parameter values and bounds that are required to best fit those measurements.

\subsection{Results}
We ran Markov Chain Monte Carlo scans using the code Monte Python 3.5~\cite{Brinckmann:2018cvx} to sample the posterior distribution of the atomic dark matter parameters given the various combinations of datasets outlined in Section~\ref{s.datasets}. From those scans, we extracted contours of 68\% and 95\% confidence level, showing the preferred regions of parameter space. We also report the best-fit and mean values, along with minimum $\chi^{2}$ values for various parameters when relevant. Because the atomic dark matter model reduces to $\Lambda\rm{CDM}+\Delta N_{\rm{eff}}$ in the limits $\fd \rightarrow 0$, $B_D \to \infty$,
and reduces to $\Lambda\rm{CDM}$ if in addition $\deltaN \rightarrow 0$, two-dimensional projections of scans over all five parameters are largely uninformative when $\Lambda\rm{CDM}$ fits the data well. Almost any choice of two out of the five parameters will appear allowed, since at least one of the parameters being marginalized over will include the $\Lambda\rm{CDM}$-like limit. We can nonetheless obtain robust constraints on $r_{\text{DAO}}$.
While we use five-dimensional scans to demonstrate how well aDM can address the $H_0$ and $S_8$ tensions simultaneously, we use three-dimensional scans with two aDM model parameters held fixed at a time to illuminate the features of the constraint contours and how they depend on the microphysical and cosmological aDM parameters.

\subsubsection{Planck + BAO + Pantheon}
\label{s.planckbaopantheon}
We first show the results of full 5-dimensional scans using the baseline datasets. 
In all scans we use use flat priors on the six standard $\Lambda\rm{CDM}$ cosmological parameters \\
$\{\omega_{b}, \omega_{dm}, h, \ln(10^{10}A_{s}), n_{s}, \tau_{\rm{reio}}\}$. 
For the aDM parameters, we use linear priors on $\fd\in [0,1]$, $\deltaN \in [10^{-4},1]$, and $\alphad \in [0.005,0.1]$.\footnote{The lower bound on $\alphad$ avoids numerical issues in CLASS due to very late dark recombination. We leave detailed exploration of this extremely weakly coupled regime for future work.} We use log priors on $\mpd \in [1,1000] \gev$ and $\med \in [0.02,100] \mev$. These bounds were chosen to avoid the regimes where basic assumptions about the thermal history of the atomic dark matter break down, as described in \ref{s.class}. These parameter ranges allow the aDM sector to vary from acting completely CDM-like to having large impacts on the cosmological history due to DAO, with the redshift of dark recombination able to vary from $z\approx 10 - 10^{9}$.

\begin{figure}
\centering
\includegraphics[width=0.7\textwidth]{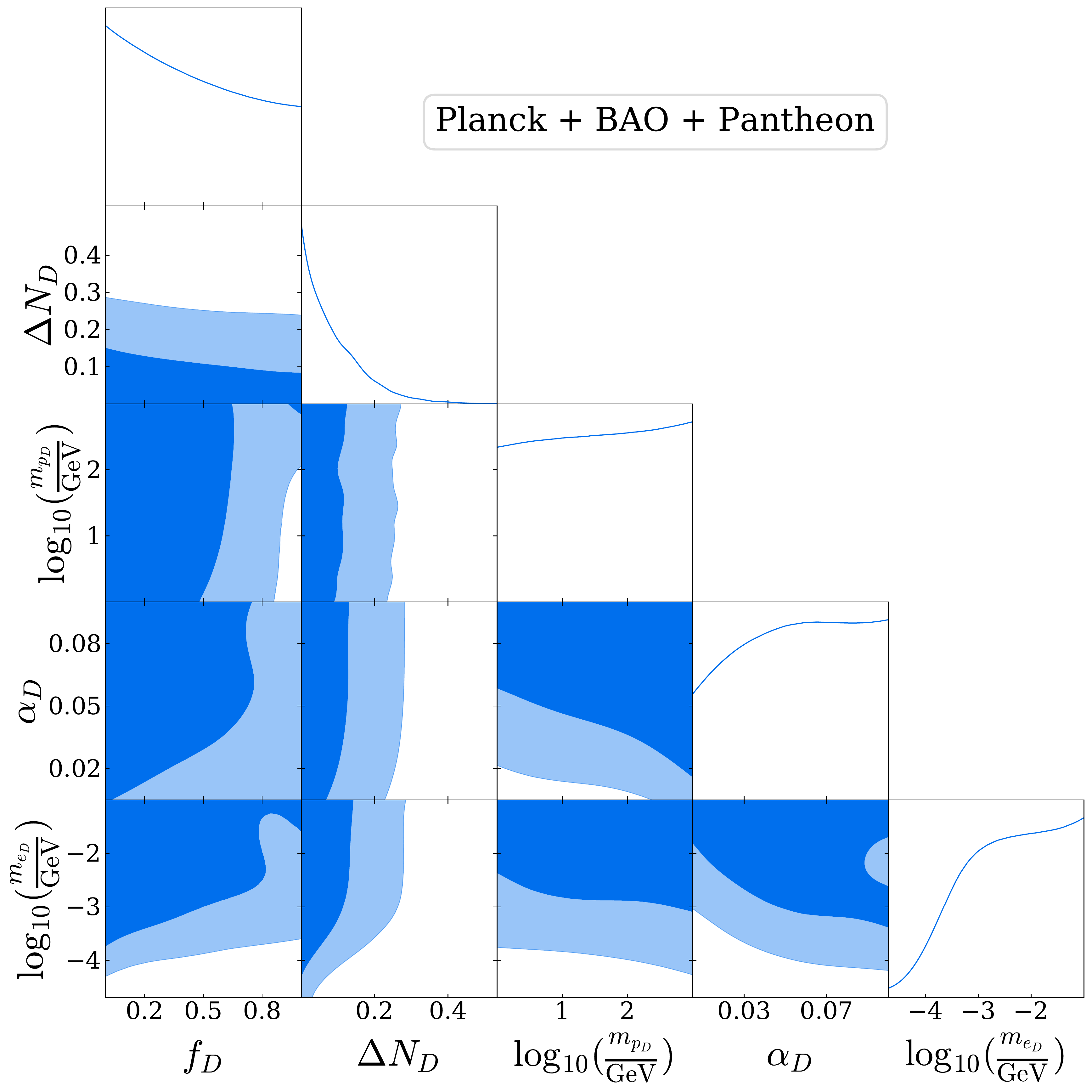}
\caption{Two-parameter projections of constraints on all five aDM parameters using the baseline dataset Planck + BAO + Pantheon. Dark and light blue show 68\% and 95\% confidence level contours. Apart from the expected $\Delta N_D \lesssim 0.3$, corresponding to $\xi \lesssim 0.5$, no combination of two parameters is significantly constrained while marginalizing over the others.}
\label{fig:PBP_allparam}
\end{figure}

\begin{figure}
\begin{center}
\begin{tabular}{ccc}

\includegraphics[width=0.45\linewidth]{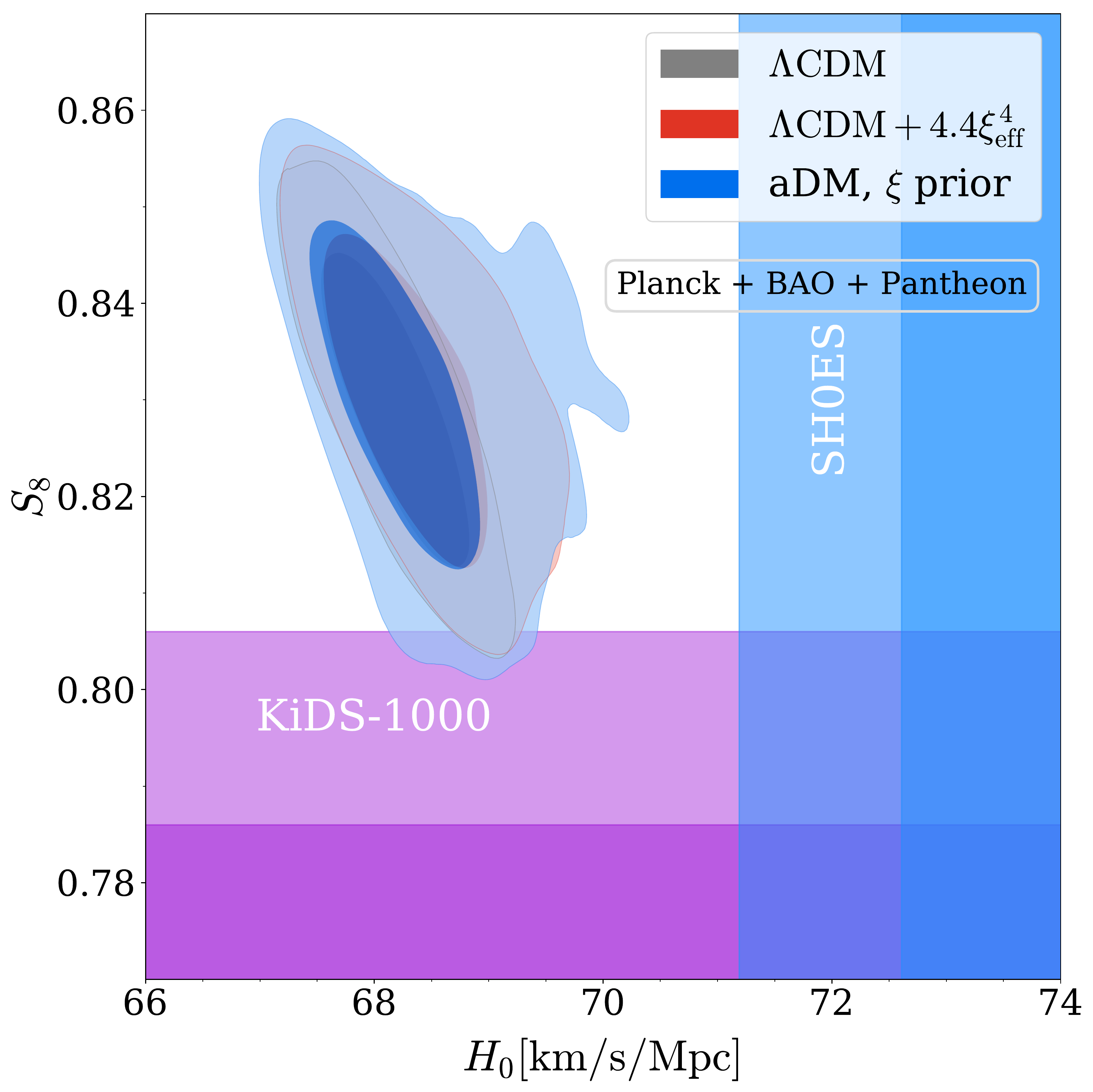}
& \phantom{bla} &
\includegraphics[width=0.45\linewidth]{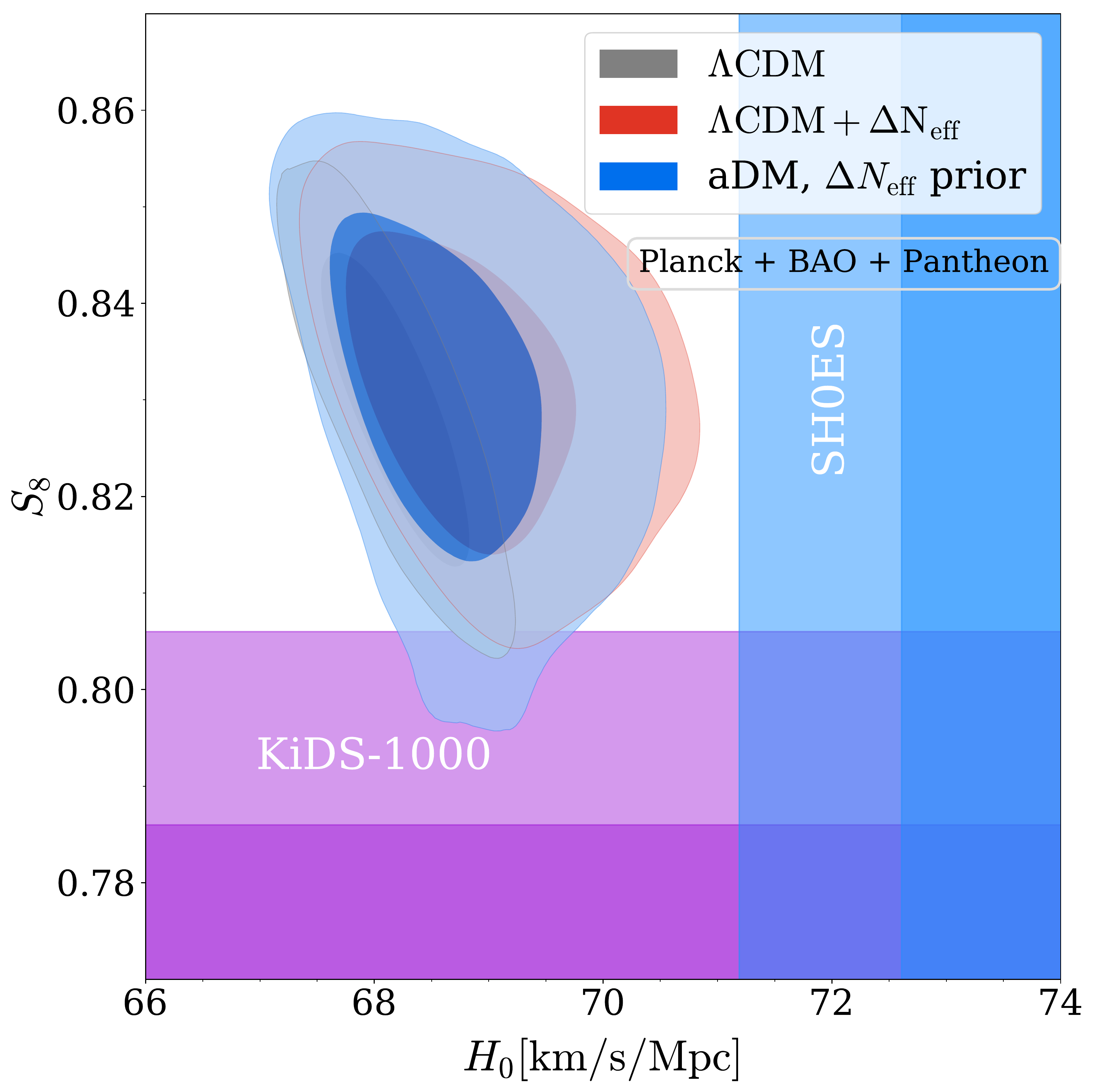}
\end{tabular}
\end{center}
\caption{
68\% and 95\% confidence level contours for $H_{0}$ and $S_{8}$ under $\Lambda\rm{CDM}$, $\Lambda\rm{CDM}+\Delta N_{\rm{eff}}$, and atomic dark matter using the baseline dataset Planck + BAO + Pantheon.
Left: Uniform prior for $\xi$, $\xi_{\rm{eff}}$.
Right: Uniform prior for $\deltaN$, $\Delta N_{\rm{eff}}$. 
}
\label{fig:PBP_H0S8}
\end{figure}

In \figref{fig:PBP_allparam} we show the two-dimensional projections of the posterior distribution of the fit of the aDM model to the baseline dataset. It is immediately clear that the only individual parameter of the model that can be robustly constrained by this scan is $\deltaN < 0.3$, corresponding to $\xi < 0.5$, which is expected from any model with free-streaming dark radiation to be consistent with Planck data. In addition, we can constrain $r_{\text{DAO}}$. Marginalized over all five model parameters, we obtain the bound $r_{\rm{DAO}} < 5.2 \rm{Mpc}$ at 95\% confidence.

From the $\med - \fd$ and $\alphad - \fd$ planes, we can see hints that $\fd$ near unity is less preferred for low $\med$ or $\alphad$. From the $\fd - \deltaN$ plane, we see that $\mathcal{O}(1)$ atomic dark matter fractions can lie within the 68\% confidence region for some parameter values, and that the upper bound on $\deltaN$ generally decreases as $\fd$ increases. Finally, the value of $\mpd$ appears to have little effect on constraints on the other model parameters. 
The high degree of degeneracy between model parameters in their effects on the CMB power spectrum makes it difficult to extract more detailed aDM constraints from this scan. 

While not very informative about constraints on any individual aDM parameter, the full five-dimensional scan is well-suited to examine the preferred region of $H_{0}$ and $S_{8}$ under the aDM model, and compare to both $\Lambda\rm{CDM}$ and $\Lambda\rm{CDM} + \Delta N_{\rm{eff}}$. To examine the impact of the choice of priors, we perform scans with two different parametrizations of the additional radiation for the $\Lambda\rm{CDM} + \Delta N_{\rm{eff}}$ and atomic dark matter models. First, we use a flat prior on $\deltaN$, or $\Delta N_{\rm{eff}}$. In the second scan, we scan over the temperature ratio $\xi$ with a flat prior in the range $[10^{-3},1]$. To match the fact that we scan over $\xi$ in the aDM model, we parametrize $\Delta N_{\rm{eff}} = 4.403\xi_{\rm{eff}}^{4}$ and scan over $\xi_{\rm{eff}}$ with a flat prior from 0 to 1. Which choice one deems more natural depends on theory bias. If the temperature asymmetry is generated through the preferential decays of a heavy right-handed neutrino to the Standard Model as in the $\nu$MTH~\cite{Chacko:2016hvu}, then $\Delta N_{\rm{eff}}$ scales with the branching ratio of the heavy particle to the dark sector, suggesting that a uniform prior on $\Delta N_{\rm{eff}}$ may be a natural choice. On the other hand, in an asymmetrically reheated Mirror Twin Higgs with a scalar reheaton, the dependence of $\Delta N_{\rm{eff}}$ on the branching ratios of the particle generating the asymmetry is highly non-trivial, and $\Delta N_{\rm{eff}}$ values spanning several orders of magnitude are realizable depending on model parameters~\cite{Craig:2016lyx,Curtin:2021alk}. It is therefore also reasonable to use a prior that explores the low-$\Delta N_{\rm{eff}}$ regime more thoroughly. 

\figref{fig:PBP_H0S8} shows the marginalized posteriors for $\lcdm$, $\lcdm + \Delta N_{\rm{eff}}$, and aDM in the $H_{0}\text{-} S_{8}$ plane, along with the $1$- and $2\text{-}\sigma$ preferred bands of $H_{0}$ from SH0ES and $S_{8}$ from the KiDS-1000 joint analysis. The left plot uses the uniform prior on $\xi$, while the right plot uses a uniform prior on $\deltaN$. Unsurprisingly, the allowed region is smaller for the scan with flat prior on $\xi$ than $\deltaN$, due to the larger volume weighting of $\xi$ values that correspond to extremely small $\Delta N_{\rm{eff}}$, and therefore smaller $H_{0}$. The minimum $\chi^{2}$ for the three models are all extremely close in value, approximately equal to $\chi^{2}_{min} = 3810$. 
We also see that the aDM contours line up closely with those of $\Lambda\rm{CDM}+\Delta N_{\rm{eff}}$, with a hint that slightly smaller $S_{8}$ values are also allowed for $H_{0}\sim 69$ km/s/Mpc. This indicates that the fit to these data does not have a strong preference for a non-zero amount of aDM, and that only the dark radiation component of the aDM model is preferred, to a small extent. As expected, both the $\lcdm{} + \Delta N_{\rm{eff}}$ and aDM best-fit regions contain the $\lcdm{}$ region, since they both can reduce to $\lcdm{}$ in appropriate parameter limits. 
By allowing extra radiation, aDM and $\lcdm{} + \Delta N_{\rm{eff}}$ can reach higher $H_{0}$ values than $\lcdm{}$, but the 95\% confidence regions of all three models are in severe tension with the local measurements of $H_{0}$ and $S_{8}$. We will later show how including local measurements of $H_{0}$ and $S_{8}$ affect these regions, and the goodness of fit of each model.

\begin{figure}
\begin{center}
\begin{tabular}{ccc}
\includegraphics[width=0.45\linewidth]{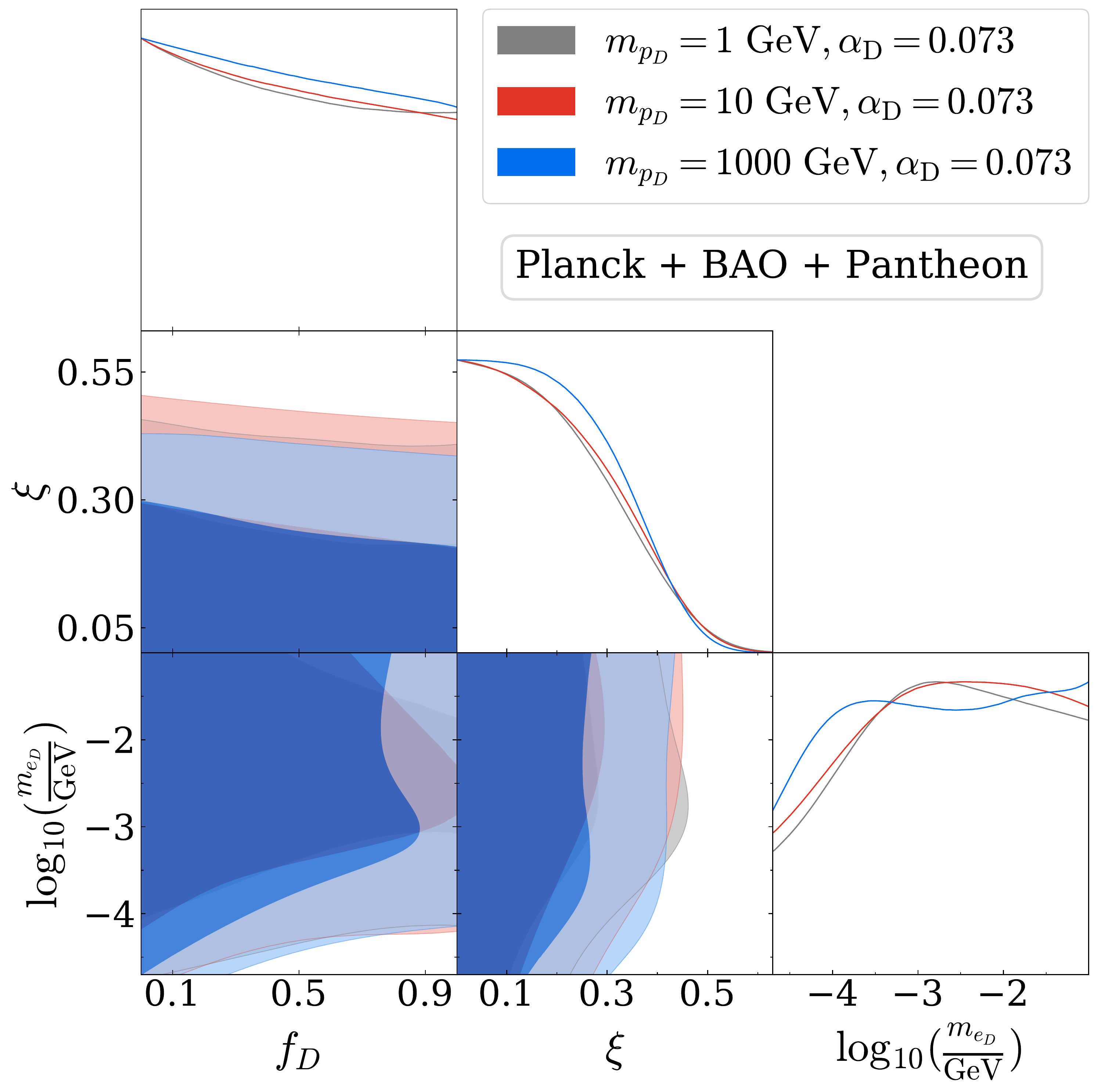}
& \phantom{bla} &
\includegraphics[width=0.45\linewidth]{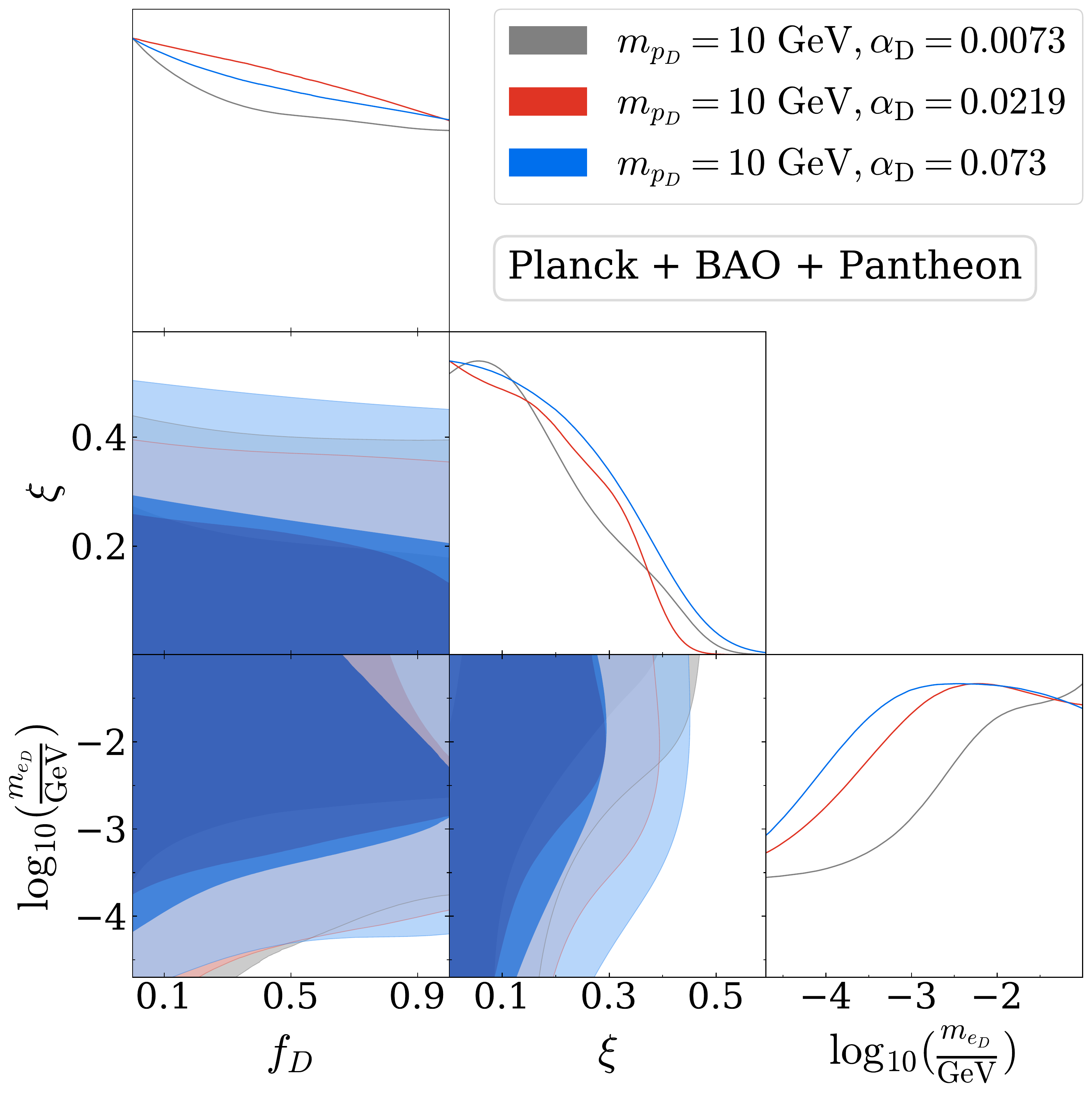}
\end{tabular}
\end{center}
\caption{
68\% and 95\% confidence level contours on aDM parameters with the baseline dataset,  for fixed values of $\mpd$ and $\alphad$. 
Left: For dark proton masses from 0.1 to 1000 GeV for $\alphad = 0.073$, demonstrating that constraints depend very little on $\mpd$.
Right: For $\alphad = 0.0073, 0.073$ with $\mpd = 10 \gev$.
}
\label{f.3dscans}
\end{figure}

\begin{figure}
\begin{center}
\includegraphics[width=0.7\textwidth]{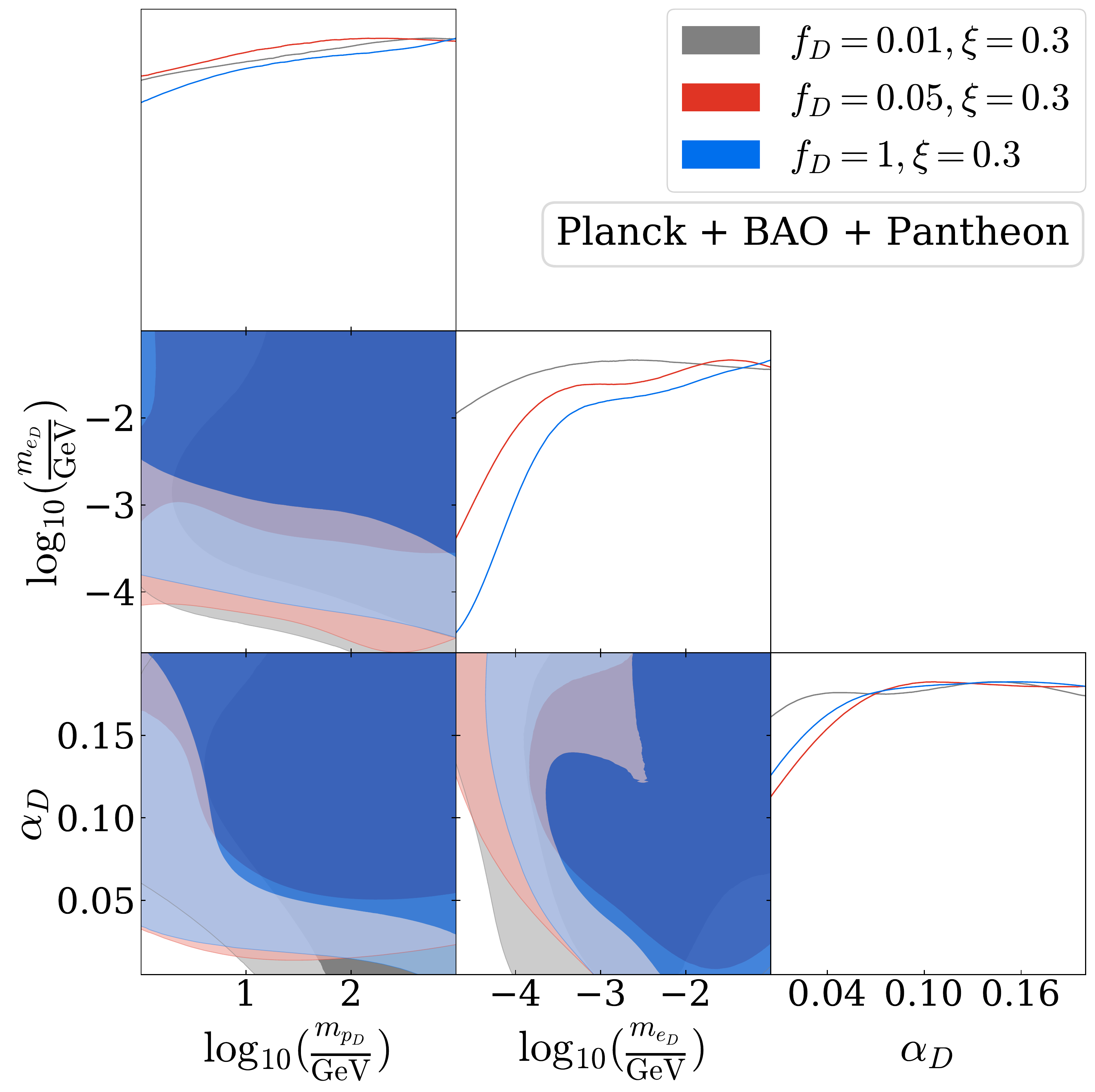}
\end{center}
\caption{68\% and 95\% confidence level contours on aDM parameters with the baseline dataset Planck + BAO + Pantheon, for fixed values of $\fd$ and $\xi$.}
\label{f.3dscans_2}
\end{figure}

\begin{figure}
\begin{center}
\includegraphics[width=0.7\textwidth]{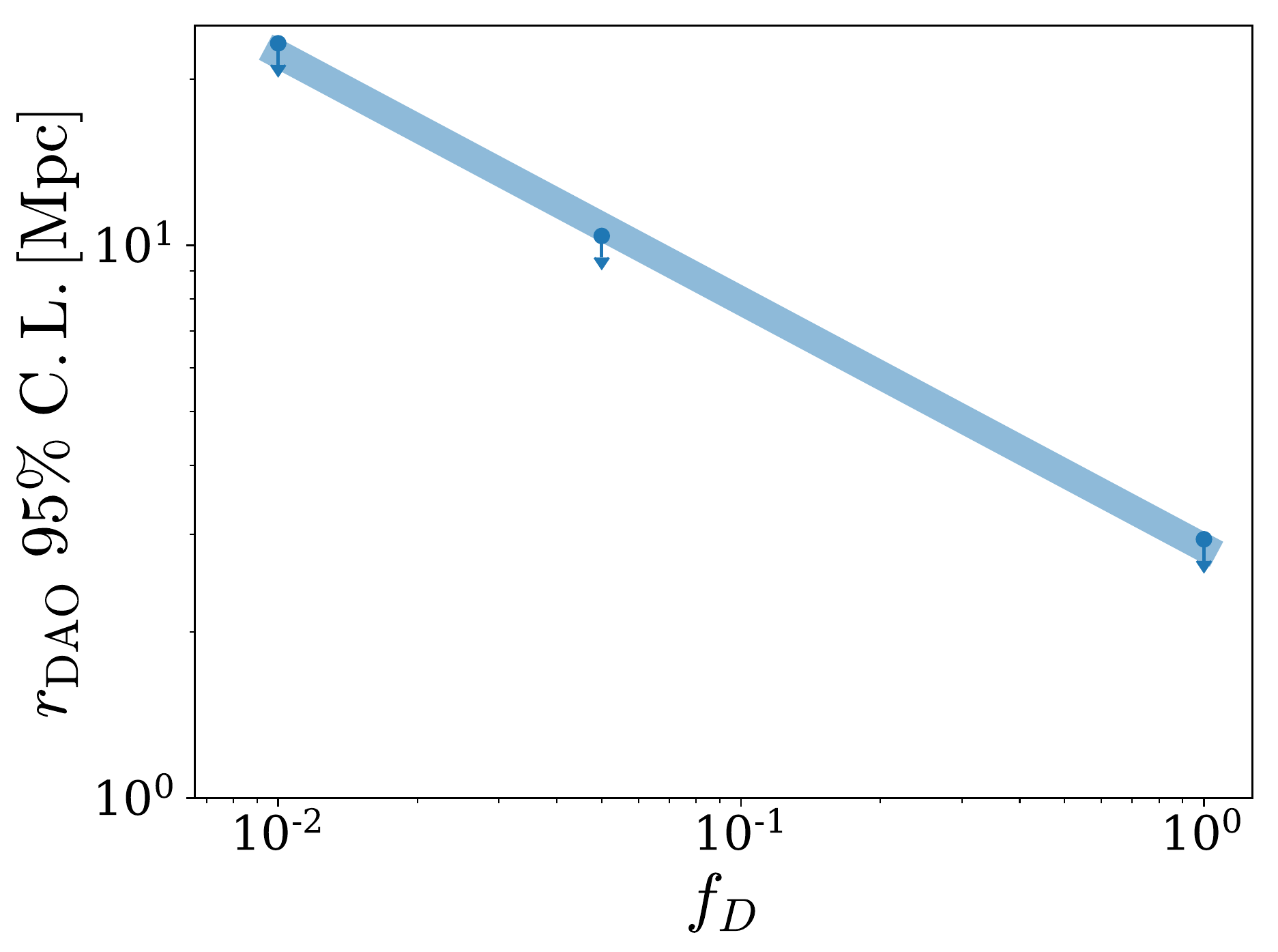}
\end{center}
\caption{Upper limits on $r_{\rm{DAO}}$ at 95\% confidence level with the baseline dataset Planck + BAO + Pantheon. Limits were found for three values of $f_{D}$, with $\xi = 0.3$, equivalent to $\Delta N_{\rm{eff}} = 0.036$. The linear fit is only to illustrate the trend of the limit as a function of $\fd$.}
\label{f.rdao_limits}
\end{figure}

To show the nature of the constraints on the total amount and temperature of the atomic dark matter, we show in Figure~\ref{f.3dscans} the results of scans holding the dark proton mass and dark fine structure constant fixed. The dark electron mass, which controls the epoch of dark decoupling, is also allowed to vary. Here the priors on $\fd$ and $\xi$ are linear priors, and $\log_{10}(m_{e_{D}}/\gev) \in [10^{-1.7},10^{2}] \mev$ is again sampled with log priors. We use a flat prior on $\xi$ instead of $\deltaN$ here to show how the bounds on $\xi$ vary with the dark QED coupling. This also clarifies the behavior of the bounds at low temperature.  
Fixing microphysical parameters (rather than cosmological parameters $f_D$, $\xi$) means that the $\Lambda$CDM limit is included in each of these 3D scans, meaning the 68\% and 95\% contours can be regarded as true constraints under the assumption that the best-fit points should fit the data at least as well as $\Lambda$CDM.

Figure~\ref{f.3dscans} (left) shows constraints for three very different fixed values of $\mpd$, while          
Figure~\ref{f.3dscans} (right)  considers three different values of $\alphad$, corresponding to 1, 3, and 10 times SM-like dark QED coupling.
We see that as $\med$ increases, $\xi$ and $f_D$ transition sharply from being extremely constrained to saturating the Planck bound on $\Delta N_{\rm eff}<0.3$ and allowing unity aDM fractions. 
The value of $\mpd$ has almost no impact on the allowed regions for the free parameters of the scan, so we can focus on the $\alphad$ dependence, which is significant.
Higher $\alphad$, corresponding to higher $B_{D}$ and therefore earlier dark recombination, allows for a larger range (i.e. smaller lower bound) of dark electron masses where bounds on $\fd, ~\xi$ are weak and dominated only by dark radiation. 
The 95\% C.L. upper limit on $r_{\rm{DAO}}$ in these scans has little dependence on $\mpd$, varying from $\sim 3$ to $\sim 4$ Mpc as $\mpd$ varies from 1000 down to 1 GeV. As $\alphad$ increases from $0.0073$ to $0.073$, the upper bound on $r_{\rm{DAO}}$ decreases from 6.5 Mpc to 3.2 Mpc. This dependence is weaker than linear, indicating that the bound on the DAO scale is fairly insensitive to the microphysics parameters. 

The dependence of the allowed $\med$ values on $\alphad$ highlights that CMB data are mostly sensitive to the sound horizon of the dark plasma at the time when the dark photons decouple from the dark protons and electrons, or equivalently the scales at which dark acoustic oscillations can impact the growth of density perturbations in the dark and visible sectors. For early enough decoupling, the dark acoustic oscillations do not impact scales observable in the CMB.

\begin{figure}
\centering
\includegraphics[width=0.7\textwidth]{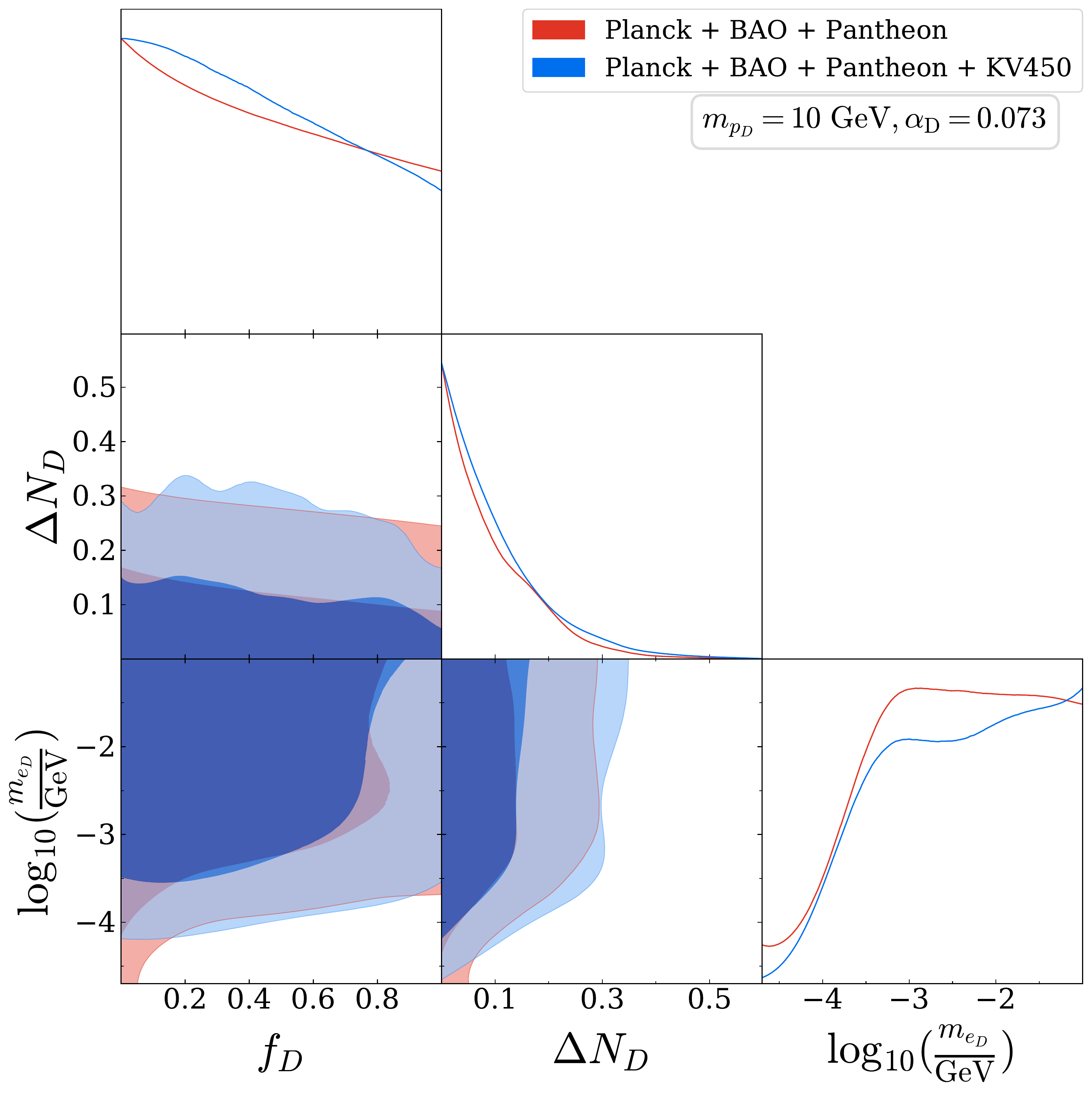}
\caption{Constraints on $\fd$, $\deltaN$, and $\med$ for fixed $\mpd = 10 \gev$ and $\alphad = 0.073$, computed with and without the KV450 cosmic shear dataset. The confidence level contours change little when the KV450 data is included, likely due to our exclusion of the non-linear regime.}
\label{fig:KV450_withvswithout}
\end{figure} 

We also performed scans with fixed $\fd$ and $\xi$ to show how $\alphad$ and $\med$ are constrained in combination as well as the upper bound on $r_{\rm{DAO}}$ for specific $\fd$ values of interest. While the $\Lambda$CDM limit is not realized in these scans, we find that for $\xi=0.3$, corresponding to $\Delta N_{\rm{eff}}=0.036$, the minimum $\chi^{2}$ is within $\approx 1$ of the minimum $\chi^{2}$ for $\lcdm$. We therefore take the derived constraints at face value. 
Figure~\ref{f.3dscans_2} shows the resulting 2D constraint contours. The constraints on $r_{\rm{DAO}}$ are as follows: For $\fd=0.01$, $\xi=0.3$, $r_{\rm{DAO}}< 23.2$ Mpc at 95\% confidence. For $\fd=0.05$, $\xi=0.3$, $r_{\rm{DAO}}< 10.4$ Mpc at 95\% confidence. For $\fd=1$, $\xi=0.3$, $r_{\rm{DAO}}< 2.9$ Mpc. These limits are shown in~\figref{f.rdao_limits}.

\subsubsection{Planck + BAO + Pantheon + KV450}
\label{s.planckbaopantheonkv450}
Through DAOs, atomic dark matter can impact not just the CMB but also the matter power spectrum, and therefore measurements of large scale structure in the universe. Because there is currently no fast, reliable calculation of the non-linear evolution of aDM perturbations, it is unwise to compare the predictions of the linear matter power spectrum to observations in the non-linear regime, $k \gtrsim 0.1 h/$Mpc. Here we show the result of including the cosmic shear tomography dataset from KiDS+Viking 450~\cite{Hildebrandt:2018yau} along with the baseline datasets in our fit, including data only up to $k=0.2h/$Mpc to limit exposure to the non-linear regime. Because of this limitation, the impact on the constraints we derive is small, as exemplified by \figref{fig:KV450_withvswithout}, where we show the result of a three-dimensional scan over $\fd$, $\deltaN$, and $\med$. The constraints on all three parameters are almost identical with or without the KV450 data.

\subsubsection{Planck + BAO + Pantheon + SH0ES + KiDS-1000}
\label{s.planckbaopantheonshoeskids}

\begin{figure}
\centering
\includegraphics[width=0.7\textwidth]{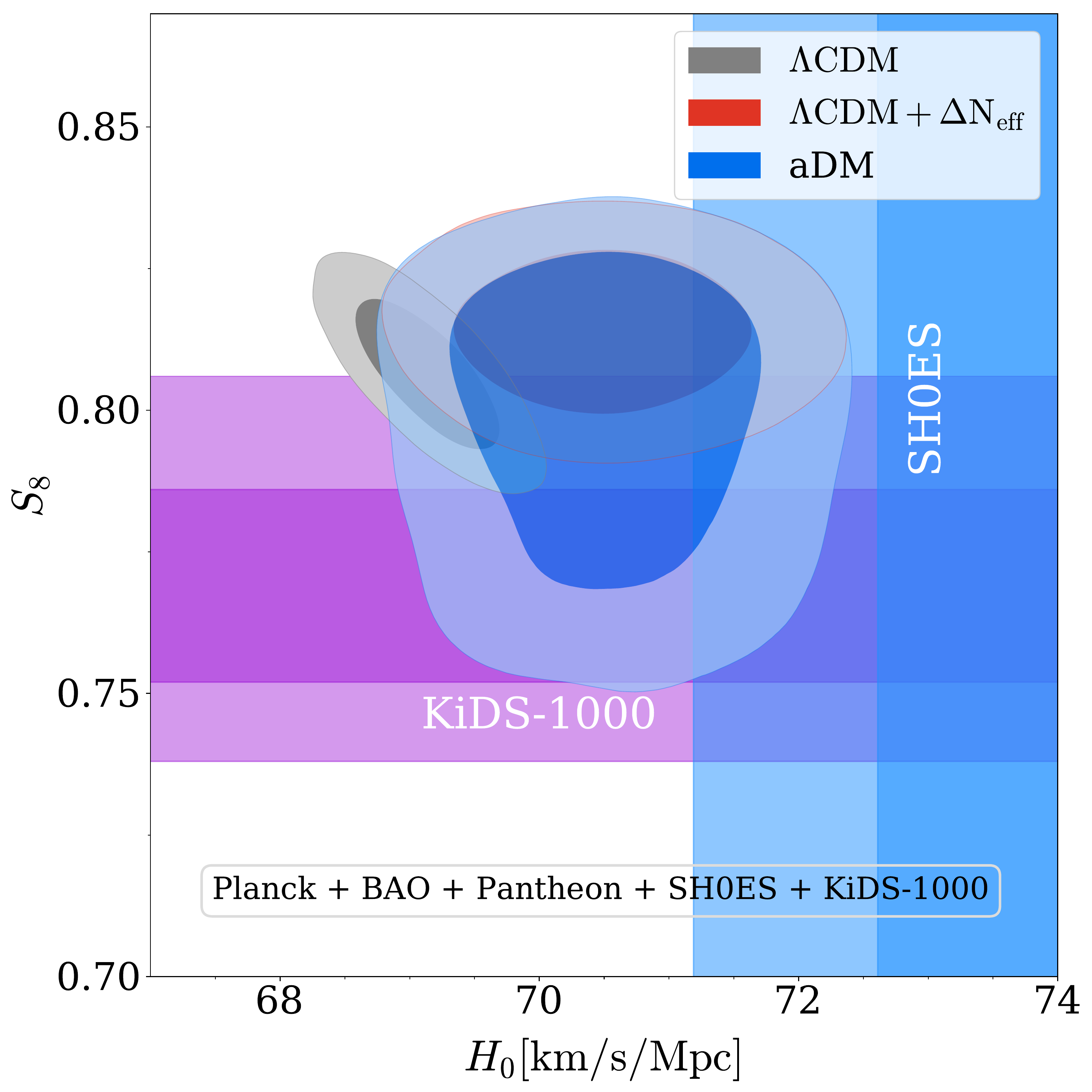}
\caption{68\% and 95\% confidence level contours for $\Lambda\rm{CDM}$, $\Lambda\rm{CDM}+\Delta N_{\rm{eff}}$, and atomic dark matter using Planck, BAO, Pantheon, SH0ES, and the KiDS-1000 measurement of $S_{8}$, demonstrating the extent to which aDM can resolve the $(H_0, S_8)$-tension.
}
\label{fig:PBPSP_H0S8}
\end{figure}

\begin{table}
\renewcommand*{\arraystretch}{1.4}
\vspace*{-9mm}
\hspace*{-9mm}
\begin{tabular}{|c|c|c|c|c|c|c|} 
\hline
& \multicolumn{2}{c|}{$\Lambda$CDM}& \multicolumn{2}{c|}{$\Lambda$CDM + $\Delta N_{\rm{eff}}$} & \multicolumn{2}{c|}{aDM}\\ 
\hline
Param        & best-fit & mean$\pm\sigma$ & best-fit & mean$\pm\sigma$ & best-fit & mean$\pm\sigma$\\
\hline 
$100~\Omega_b h^2$ &$2.266$ & $2.261_{-0.013}^{+0.013}$ &$2.273$ & $2.275_{-0.015}^{+0.015}$ &$2.280$ & $2.275_{-0.014}^{+0.014}$  \\ 
$\Omega_{dm } h^2$ & $0.1173$ & $0.1174_{-0.0008}^{+0.0008}$  &$0.1207$ & $0.1217_{-0.0024}^{+0.0022}$ &$0.1238$ & $0.1228_{-0.0028}^{+0.0024}$ \\ 
$h$ & $0.6924$ & $0.691_{-0.0037}^{+0.0037}$ &$0.7056$ & $0.7053_{-0.0080}^{+0.0072}$&$0.7077$ & $0.7055_{-0.0078}^{+0.0078}$  \\ 
$\ln(10^{10}A_{s })$  & $3.043$ & $3.043_{-0.015}^{+0.014}$ &$3.055$ & $3.051_{-0.016}^{+0.014}$&$3.057$ & $3.054_{-0.014}^{+0.014}$ \\ 
$n_{s }$      & $0.9720$ & $0.9713_{-0.0037}^{+0.0036}$&$0.9763$ & $0.9789_{-0.0051}^{+0.0049}$ &$0.9737$ & $0.9765_{-0.0057}^{+0.0050}$   \\ 
$\tau{}_{reio }$  &$0.05663$ & $0.05629_{-0.0077}^{+0.0069}$&$.05805$ & $0.05537_{-0.0075}^{+0.0071}$ &$0.05713$ & $0.05665_{-0.0072}^{+0.0072}$ \\ 
$H_0$ [km/s/Mpc]        &$69.24$ & $69.10_{-0.37}^{+0.37}$ &$70.56$ & $70.53_{-0.80}^{+0.72}$ &$70.77$ & $70.55_{-0.78}^{+0.78}$ \\ 
$\sigma_8$     & $0.8159$ & $0.8162_{-0.0059}^{+0.0057}$  &$0.8256$ & $0.8274_{-0.0082}^{+0.0078}$&$0.7832$ & $0.8096^{+0.027}_{-0.014}$\\ 
$S_8$     & $0.8049$ & $0.8065_{-0.0087}^{+0.0087}$ &$0.8091$ & $0.8140_{-0.0095}^{+0.0095}$  &$0.7737$ & $0.799_{-0.015}^{+0.024}$  \\ 
\hline
$\fd$      &- &- &- & -&$0.63$ & $<0.62$ \\ 
$\Delta N_{\rm{eff}}$    &- & -&$0.27$ & $0.25_{-0.13}^{+0.12}$ &$0.360$ & $0.296_{-0.14}^{+0.14}$   \\ 
$\log_{10}(\mpd/\gev)$     &- &- & -&-&$2.15$ & $1.55$\\
$\log_{10}(\med/\gev$) &- & -&- &- &$-3.4$ & $-3.0_{-1.1}^{+0.6}$ \\
$\alphad$ &- & -&- &- &$0.032$ & $0.054$ \\
\hline 
$\mathbf{\chi^2_{total}}$  & \multicolumn{2}{c|}{3833.87} & \multicolumn{2}{c|}{3831.11} & \multicolumn{2}{c|}{3827.23}\\
\hline \hline
Planck         & \multicolumn{2}{c|}{2774.26} & \multicolumn{2}{c|}{2777.20} & \multicolumn{2}{c|}{2780.05}\\
Pantheon & \multicolumn{2}{c|}{1025.85} & \multicolumn{2}{c|}{1025.87} & \multicolumn{2}{c|}{1026.21} \\
BAO           & \multicolumn{2}{c|}{7.24}   & \multicolumn{2}{c|}{8.448}   & \multicolumn{2}{c|}{6.77}\\
SH0ES          & \multicolumn{2}{c|}{13.36}  & \multicolumn{2}{c|}{5.69}   & \multicolumn{2}{c|}{4.76}\\		
Lensing         & \multicolumn{2}{c|}{9.40}  & \multicolumn{2}{c|}{9.26}  & \multicolumn{2}{c|}{9.29}\\		
KiDS-1000 $S_{8}$        & \multicolumn{2}{c|}{3.78}  & \multicolumn{2}{c|}{4.65}  & \multicolumn{2}{c|}{0.147}\\		
\hline 
\end{tabular} 

\caption{The mean and best-fit values for the $\lcdm{}$, $\lcdm{} + \Delta N_{\rm{eff}}$ and aDM models obtained using the Planck, BAO, Pantheon, SH0ES, and KiDS-1000 datasets. For the aDM parameters, uncertainties are included if available. In the lower part of the table, total $\chi^2$ of the best-fit points of the three models, along with the breakdown of contributions from the different datasets, is shown.  $\chi^2$ for the Planck Lensing data is shown separately with the label ``Lensing,'' and is not included in the combined $\chi^2$ for Planck.}
\label{tab:PlanckBAO_H0_SZ}
\end{table}

\begin{figure}
\centering
\includegraphics[width=0.95\textwidth]{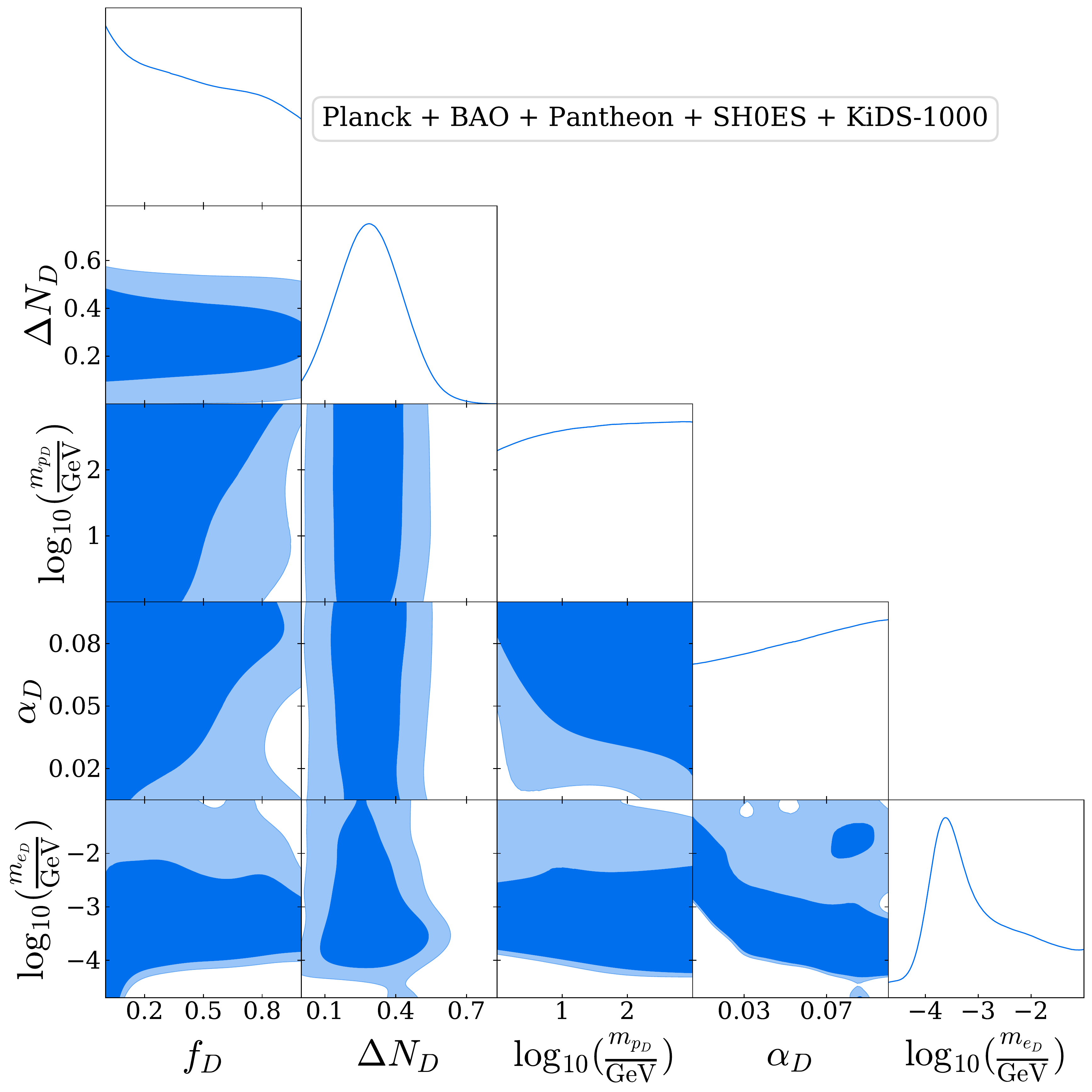}
\caption{68\% and 95\% confidence level contours for atomic dark matter model parameters using Planck, BAO, Pantheon, SH0ES, and the KiDS-1000 measurement of $S_{8}$. }
\label{fig:PBPSP_allparam}
\end{figure}   

To assess how much atomic dark matter can relieve the $H_{0}$ and $S_{8}$ tensions simultaneously relative to $\lcdm{}$ with and without dark radiation, we sample the posterior for each model including the baseline dataset, the SH0ES measurement of $H_{0}$, and the KiDS-1000 measurement of $S_{8}$. These last two likelihoods are implemented as asymmetric Gaussians using the quoted best-fit values and uncertainties. The priors for the $\lcdm{}$ and atomic dark matter parameters are the same as for the scans using only the baseline dataset. The two-dimensional marginalized posterior in the $H_{0}-S_{8}$ plane is displayed in \figref{fig:PBPSP_H0S8}. While the best-fit regions for all three models are pulled to higher $H_{0}$ and lower $S_{8}$ than with only the baseline dataset, the aDM  95\% confidence region is larger and pulled much further than both $\lcdm{}$ and $\lcdm{} + \Delta N_{\rm{eff}}$. In particular, the $\lcdm{} + \Delta N_{\rm{eff}}$ region is pulled to higher $H_{0}$ than $\lcdm{}$, but not lower $S_{8}$, demonstrating that free-streaming dark radiation alone cannot resolve both tensions at once. It is the interactions between the atomic dark matter and dark radiation that allow the model to fit a lower $S_{8}$. 

\begin{figure}
\centering
\includegraphics[width=0.7\textwidth]{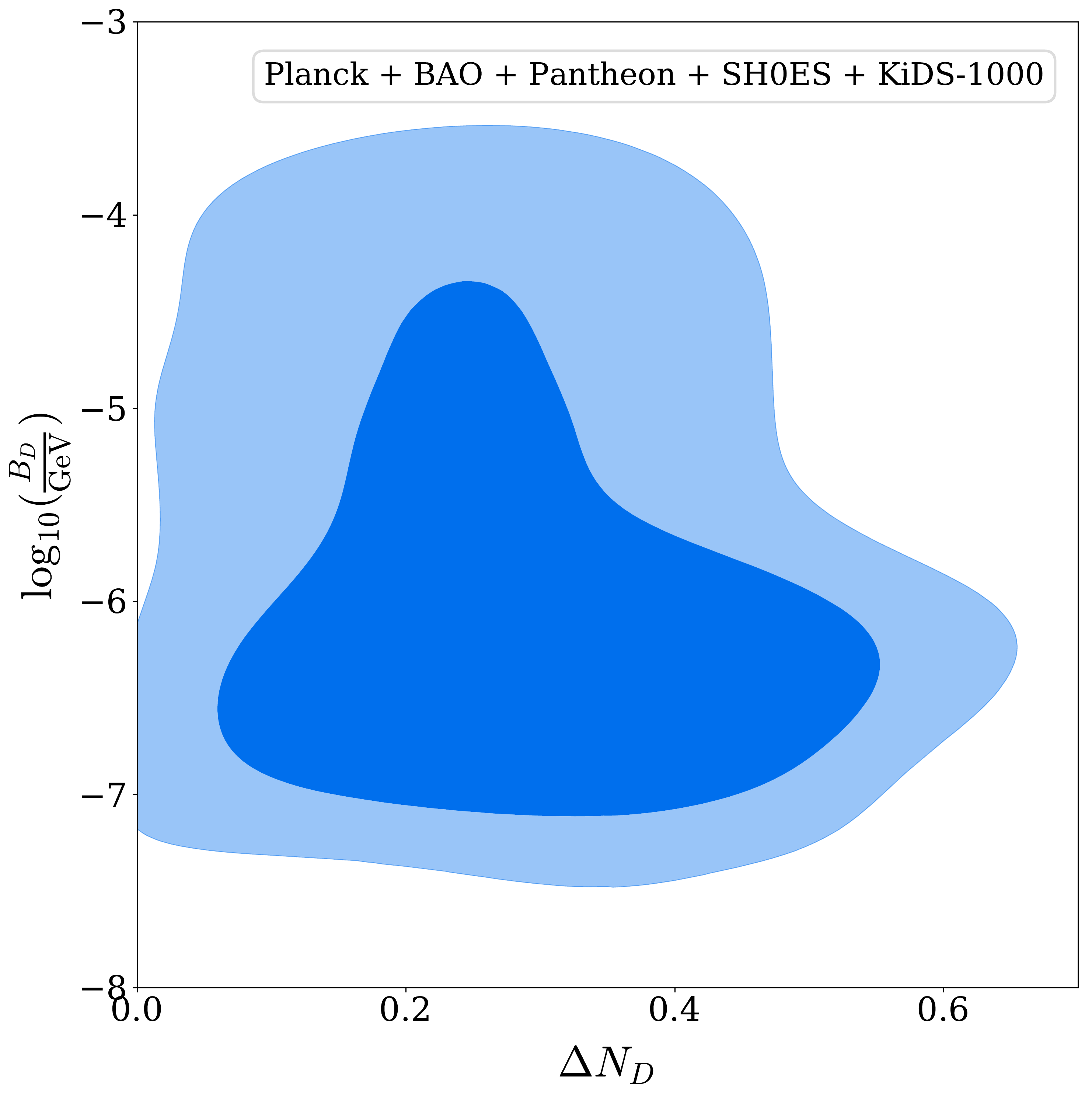}
\caption{68\% and 95\% confidence level contours for $\deltaN$ and $B_{D}$ using Planck, BAO, Pantheon, SH0ES, and the KiDS-1000 measurement of $S_{8}$. }
\label{fig:PBPSP_deltaN_BD}
\end{figure}

We summarize the best-fit and mean cosmological and aDM model parameters in Table~\ref{tab:PlanckBAO_H0_SZ}, as well as the minimum $\chi^{2}$ values for each model, and breakdown by experiment. The minimum $\chi^{2}$ for the aDM model is lower than for $\lcdm$ and $\lcdm + \Delta N_{\rm{eff}}$, but not enough to claim a significant preference for the model. Figure \figref{fig:PBPSP_allparam} shows the constraints on the aDM model parameters with the local measurements included. We see that the 95\% confidence level contours for $\deltaN$ are now closed from below, and that there is a preference for values of $\med$ and $\alphad$ that lie along a particular contour, corresponding to a best-fit $B_{D}\approx 10^{2.3} \ev$. The mean and 1-$\sigma$ range for $\log_{10}(B_{D})$ is $-5.96^{+0.39}_{-1.1}\ \gev$. Figure \figref{fig:PBPSP_deltaN_BD} shows the constraints in the space of $\deltaN$ and $B_{D}$. 

A non-zero amount of dark radiation is clearly preferred by the local measurements, and a particular dark binding energy is picked out, but there is no clear preference for a non-zero amount of atomic dark matter. While not preferred by CMB and BAO data alone, an atomic dark matter sector can reduce the $H_{0}$ and $S_{8}$ tensions by significantly broadening the range of $H_{0}$ and $S_{8}$ values that fit the data well within the model. It is interesting to note that the required range of $\deltaN$ lies exactly in the natural range given expected entropy injections in the visible sector after SM-aDM decoupling. 
    
\section{Conclusion}

\label{sec:conclusions}

Atomic dark matter is a simple theory for a dark sector that could account for some or all of our universe's dark matter abundance. 
It is highly theoretically motivated, since it could arise in dark sectors related to the SM by some discrete symmetry, which includes the Twin Higgs family of solutions to the hierarchy problem. 
The presence of dark radiation and dark acoustic oscillations also motivate aDM as a candidate to resolve the $(H_0, S_8)$ tension.
Indeed, the self-interacting and dissipative dynamics of any aDM subcomponent leads to rich phenomenology at all scales, from early universe cosmology to stellar astrophysics. 
However, the same richness also leads to great difficulty in relating constraints from smaller scales to cosmological parameters of aDM. That makes robust bounds  from precision cosmology in the linear regime all the more important. 
In this work, we derive such bounds on the full 5D aDM parameter space of three microphysical parameters $(\alphad, \mpd, \med)$ plus the aDM fraction $\fd$ and the temperature ratio $\xi$ (or equivalently $\Delta N_D$) for the first time. 

Without late-time measurements, considering only Planck + BAO + Pantheon data, we find relatively modest but still significant constraints: If 1\% of dark matter is atomic, the DAO scale is constrained to be $r_{\rm{DAO}}<23.3$ Mpc, but if $r_{\rm{DAO}}\lesssim 3$ Mpc, near-unity aDM fractions are allowed as long as the dark radiation does not violate $\Delta N_D \lesssim 0.3$. 
Adding large scale structure data in the linear regime by including the KV450 dataset has negligible impact, which motivates understanding non-linear structure growth in the presence of an aDM component. 
Including late-time measurements shows that aDM  can accommodate the $(H_0, S_8)$ tension better than $\Lambda$CDM. Points in the aDM parameter space that result in $H_{0}$ and $S_{8}$ values closer to the local measurements pick out a dark binding energy $\log_{10}(B_{D}/
\mathrm{GeV})\approx -5.96^{+0.39}_{-1.1}$.

While we studied the minimal aDM model in near-full generality, there are several important avenues for future investigation even within the specific scope of precision cosmology. For example, it would be interesting to numerically investigate the regime of very small $\alphad$ to see if a lower bound on the dark QED coupling can be found in some circumstances. Extending the CLASS-aDM code to include Case-A recombination would enable the exploration of the more weakly coupled regime. Non-minimal aDM scenarios, beyond the specific Mirror Twin Higgs realization studied in~\cite{Bansal:2021dfh}, should also be considered. We leave this for future work. 

The enormous range of astrophysical phenomena that could be realized by aDM, as well as the great difficulty of understanding them in detail, has in the past stymied detailed investigation of its signatures at non-linear or astrophysical scales and their connection to primordial parameters $(f_D, \Delta N_D)$.
Our results therefore serve as an important new jumping-off point for the rigorous study of atomic dark matter in our universe at all scales. 

Late-time $(H_0, S_8)$ measurements shine a light on a particular region of aDM parameter space that has to be the target of detailed simulation studies to push our understanding of aDM from the early universe to later times, making contact with treasure troves of data along the way.
This could ultimately lead to the discovery of non-minimal dark sectors even in the complete absence of non-gravitational interactions with the SM. With sufficient study of its detailed evolution, distribution and gravitational effects, the nightmare scenario of the perpetually unknown dark matter that minimally interacts with the SM can thus be avoided. 

\subsection*{Acknowledgements}
We are especially grateful to Francis-Yan Cyr-Racine for many helpful conversations on aDM cosmology;
Yacine Ali-Ha\"{i}moud and Nanoom Lee for useful advice and providing the code to compute the effective recombination coefficients in HyRec;
Melissa Joseph and Kylar Greene for technical advice on MCMC scans;
and Zackaria Chacko for initial encouragement for this project.
We also thank 
Daniel Gruner and Bruno Mundim for technical assistance on the Niagara cluster.
This work was enabled by computational resources provided by Compute Canada and the Digital Research Alliance of Canada.
The research of SB in part was supported by the DOE grant DE-SC0011784.
The research of JB and DC was supported in part by a Discovery Grant from the Natural Sciences and Engineering Research Council of Canada, the Canada Research Chair program, the Alfred P. Sloan Foundation, the Ontario Early Researcher Award, and the University of Toronto McLean Award.
JB also acknowledges funding from a Postgraduate Doctoral Scholarship (PGS D) provided by the Natural Sciences and Engineering Research Council of Canada.
The research of YT was supported by the NSF grant  PHY-2112540.

\bibliographystyle{Jhep}
\bibliography{references}
\end{document}